\documentclass[]{aa}

\usepackage{graphicx}
\usepackage{txfonts}
\usepackage{xcolor}
\usepackage[breaklinks=true]{hyperref}
\hypersetup{
    colorlinks=true,
    citecolor=blue,
    linkcolor=blue,
    filecolor=magenta,      
    urlcolor=blue,
    }

\usepackage{natbib,twoopt}
\usepackage[breaklinks=true]{hyperref} 
\bibpunct{(}{)}{;}{a}{}{,} 
\makeatletter
\newcommandtwoopt{\citeads}[3][][]{\href{http://adsabs.harvard.edu/abs/#3}%
{\def\hyper@linkstart##1##2{}%
\let\hyper@linkend\@empty\citealp[#1][#2]{#3}}}
\newcommandtwoopt{\citepads}[3][][]{\href{http://adsabs.harvard.edu/abs/#3}%
{\def\hyper@linkstart##1##2{}%
\let\hyper@linkend\@empty\citep[#1][#2]{#3}}}
\newcommandtwoopt{\citetads}[3][][]{\href{http://adsabs.harvard.edu/abs/#3}%
{\def\hyper@linkstart##1##2{}%
\let\hyper@linkend\@empty\citet[#1][#2]{#3}}}
\newcommandtwoopt{\citeyearads}[3][][]%
{\href{http://adsabs.harvard.edu/abs/#3}
{\def\hyper@linkstart##1##2{}%
\let\hyper@linkend\@empty\citeyear[#1][#2]{#3}}}
\makeatother

\usepackage{multirow}
\begin{document}

   \title{The warm-hot intergalactic medium in inter-cluster filaments}

   \subtitle{A forecast for HUBS observations based on eRASS1 superclusters}

   \author{Yuanyuan Zhao\inst{1,2,3},
          Haiguang Xu\inst{2}\thanks{corresponding author; hgxu@sjtu.edu.cn},
          Ang Liu\inst{4},
          Xiaoyuan Zhang\inst{4},
          Li Ji\inst{5},
          Jiang Chang\inst{5},
          Dan Hu\inst{6},
          Norbert Werner\inst{6},
          Zhongli Zhang\inst{3},
          Wei Cui\inst{7},
          Xiangping Wu\inst{8}
          }

   \institute{Tsung-Dao Lee Institute, Shanghai Jiao Tong University, 1 Lisuo Road, Pudong New Area, Shanghai, 201210, China
        \and
            School of Physics and Astronomy, Shanghai Jiao Tong University, 800 Dongchuan Road, Shanghai 200240, China  \\
              \email{yuanyuan.zhao@sjtu.edu.cn; hgxu@sjtu.edu.cn}
        \and
            Shanghai Astronomical Observatory, Key Laboratory of Radio Astronomy, Chinese Academy of Sciences, Shanghai 200030, China
         \and
             Max Planck Institute for Extraterrestrial Physics, Giessenbachstrasse 1, 85748 Garching, Germany
        \and 
            Purple Mountain Observatory, Chinese Academy of Science, Nanjing 210034, China
        \and
            Department of Theoretical Physics and Astrophysics, Faculty of Science, Masaryk University, Kotl\'{a}\v{r}sk\'{a} 2, Brno, 611 37, Czech Republic
        \and
            Department of Astronomy, Tsinghua University, Beijing 100084, China
        \and
            National Astronomical Observatories, Chinese Academy of Sciences, 20A Datun Road, Beĳing 100012, China
             }

   \date{}


\abstract
   {Cosmological simulations indicate that nearly half of the baryons in the nearby Universe are in the warm-hot intergalactic medium (WHIM) phase, and about a half them reside in cosmic filaments connecting galaxy clusters. Recent observational studies using stacked survey data and deep exposures of galaxy cluster outskirts have detected soft X-ray excesses associated with optically identified filaments. However, the physical characteristics of WHIM in filaments remain largely undetermined due to a lack of direct spectral diagnostics of individual targets, which are limited by the spectral resolution of current instruments in the soft X-ray band. }
   {We aim to select appropriate targets for WHIM characterization through pointing observations with the future Hot Universe Baryon Surveyor (HUBS) mission, which is designed with eV-level energy resolution in the 0.1-2.0 keV band and a one-square-degree field of view, thus complementing other planned microcalorimetry missions such as Athena.}
   {We built a sample of 1577 inter-cluster filaments based on the first eROSITA All-Sky Survey (eRASS1) supercluster catalog and estimated their soft X-ray emission. Their modeled emission and geometrical properties were used to select candidate targets for HUBS observations.}
  {Four inter-cluster filaments were selected as the most appropriate candidates. By simulating and analyzing their mock observations, we demonstrated that with 200 ks HUBS exposure for each candidate, the gas properties of individual filaments can be accurately determined, with the temperature constrained to $\pm0.01$ keV, metallicity constrained to $\leq\pm0.03$ solar, and density constrained to $<\pm10\%$. Elemental abundances of O, Ne, Mg, and Fe can be measured separately, providing unprecedented insights into the chemical history of the filament gas. We also show that direct mapping of the WHIM distribution is promising with narrowband imaging of the O \textsc{viii} line.}
   {Our work forecasts that next-generation X-ray missions such as HUBS will provide substantial improvement in our understanding of the physical status and evolution history of the diffuse WHIM gas in the cosmic large-scale structure.}

   \keywords{large-scale structure of Universe --
                intergalactic medium --
                techniques: imaging spectroscopy -- instrumentation: detectors
               }
   \titlerunning{WHIM filaments with HUBS}
   \authorrunning{Y. Zhao et al.}
   \maketitle

%
\renewcommand{\arraystretch}{1.5}
\section{Introduction}\label{sec:intro}
In the past five decades, galaxy surveys and multi-band sky mappings \citep[e.g.,][]{York2000, Liske2015, Colless2001, Secco2022} have revealed that the nearby Universe exhibits a highly inhomogeneous, complex structure that is characterized by a multi-scale network consisting of gigantic sheets and filaments, which, separated by numerous voids, connect to high-density nodes where massive galaxy clusters are formed through mergers and accretion of smaller structures \citep{Bond1994, Davis1985, Frenk1985, Bond1996, Sarazin2002, Aragon-Calvo2010, Cautun2013}. Within the frame of the standard cosmological model, the structures in this cosmic web can be reconstructed in modern large-scale cosmological simulations, such as IllustrisTNG \citep{Nelson2019}, and EAGLE \citep{Schaye2015}, and FLAMINGO \citep{Schaye2023}, to list a few. As predicted in these simulations, the filaments have gathered about 50 \% of the mass budget in the Universe since $z = 2$, compared with about $\sim 25\%$ for sheets, $\sim 15\%$ for voids, and $\sim 10\%$ for nodes \citep{Cautun2014, Cornwell2022}. Meanwhile, it has also been predicted that a significant part of galaxies are embedded in the filaments.

Besides the baryons in galaxies, cosmological simulations suggest that in the cosmic filaments there should exist an abundance of diffuse baryons, which permeate the vast regions between galaxies. These baryons are confined in the dark matter-dominated gravitating potential well, and their total mass overwhelms that of the galaxies by a factor of about ten \citep{Peebles1975, Klypin1983, Zhang1995, Hernquist1996, Fukugita2004, Cen2006, Prochaska2009, Haider2016, Mandelker2019}. In today's Universe, it is predicted that less than about $40\%$ of these diffuse baryons remain in a relatively cool ($<10^{5}$ K) and low density ($10^{-7}-10^{-5}$ cm$^{-3}$) phase, usually referred to as the diffuse intergalactic medium (IGM). The other $40-50\%$ of the diffuse baryons are believed to have been shock heated during gravitational collapse and structure formation to reach temperatures of $10^{5}-10^{7} \textrm{K}$ (densities of $10^{-7}-10^{-5} \textrm{cm}^{-3}$; \citealt{Cen1999, Dave2001, Valageas2002, Kang2005, Cen2006, Cautun2014, Martizzi2019}), thus called the warm-hot intergalactic medium (WHIM). The results of these simulation have led to the consensus that the WHIM in the filaments constitutes the largest reservoir for $30-40\%$ of the primordial baryons that have not been detected in the nearby Universe (the ``missing baryon'' problem; \citealt{Fukugita1998, Cen1999, Soltan2002, Nicastro2005, Shull2012}).

Since the 2000s, X-ray instruments equipped with sensitive charge-coupled devices (CCD) have provided a prospective approach to detect WHIM, whose abundant characteristic emission lines can be best probed in the soft X-ray band. By analyzing the data obtained from a deep XMM-Newton observation, \cite{Werner2008} reported a wavelet-based detection of X-ray emission ($0.5-2$ keV, $5\sigma$) from a 1.2 Mpc-wide filament between the closely interacting clusters Abell 222 and Abell 223. Also using XMM-Newton data ($0.5-1.2$ keV), \cite{Eckert2015} found filamentary structures on megaparsec scales outside the viral radius (2.1 Mpc) of the massive cluster Abell 2744, which coincide with the local overdensities of galaxies and dark matter. Similarly, in Abell 1750 \cite{Bulbul2016} detected cool ($0.8-1$ keV) gas along a large-scale filament at the virial radius with Suzaku. \cite{Sarkar2022b} have also reported the detection of an inter-cluster gas filament in the Abell 98 system along the merger axis with a temperature of $1.07 \pm 0.29$ keV using Chandra observations. We note that the WHIM studied in these works all reside in the outskirt region of galaxy clusters (Abell 222 and Abell 223 have approached each other to within their virial radii), that is, in the transition zone between filament gas and cluster gas, while the main bodies of the WHIM in cosmic filaments, which can reach tens of megaparsecs in length, remained untouched. 

More recently, significant progress has been made by applying stacking techniques to X-ray data obtained with large survey telescopes such as the extended ROentgen Survey with an Imaging Telescope Array (eROSITA; \citealt{Predehl2021}) on board the Spectrum-Roentgen-Gamma (SRG) mission. \cite{Tanimura2022} stacked the 0.4 – 2.3 keV data of the eROSITA Final Equatorial Depth Survey (eFEDS) for 463 filaments identified in the Sloan Digital Sky Survey ($0.2 < z < 0.6$, $30 - 100$ Mpc in length), and detected an X-ray excess at a significance of $3.8\sigma$. The measured gas overdensity at the filament center and gas temperature averaged for filament-core regions are $\delta \sim 21$ and $T = 1.0^{+0.3}_{-0.2}$ keV, which is in agreement with the hotter phase of WHIM from cosmological simulations. Using the first four scans of the SRG/eROSITA All Sky Survey (eRASS:4) data, \cite{Zhang2024} stacked 0.3 - 1.2 keV data of 7817 optically identified filaments and achieved 9.0$\sigma$ detection of excess X-ray emission, which has best-fit temperature of log($T$/K) = 6.84 $\pm$ 0.07 and average density contrast of log$\Delta_\textrm{b} = 1.88 \pm 0.18$, and estimated 60\% of the excess X-ray emission can be attributed to WHIM after considering the contamination from unmasked galactic halos, active galactic nuclei (AGN), and X-ray binaries. In addition to these discoveries through stacking analysis, with deep eROSITA pointing observations, \cite{Reiprich2021} successfully detected a 15 Mpc (projected) X-ray filament in the Abell 3391–Abell 3395 system, as well as multiple substructures beyond the virial radius, the positions of which are consistent with the structure found in Sunyaev-Zeldovich (SZ) and optical observations. Using eRASS:4 survey data, \cite{Dietl2024} also discovered an X-ray excess beyond three times the virial radius of Abell 3667, smoothly connecting to A3651, indicating an inter-cluster filament of $25-32$ Mpc in length.

Despite the progress, challenges remain with regard to the details of the physical state and evolution history of WHIM. First, current statistical measurements of gas overdensity and gas temperature, which are largely based on stacking technique, have not been convincingly constrained to the level that can be used to fully address the ``missing baryon'' problem \citep{Driver2021}. The situation becomes even more complicated when considering (a) the difference in gas temperatures measured with X-ray data and with combined SZ and cosmic microwave background (CMB) lensing data \citep{deGraaff2019, Tanimura2020, Tanimura2022, Zhang2024}, and (b) it has been reported that physical states tend to be different for filaments with different lengths \citep[e.g.,][]{Galarraga-Espinosa2021}. Second, the details of the chemical status and enrichment history of WHIM in both cosmic filaments and the filament-cluster transition zone, which are tightly connected to the studies of the cosmic feedback cycle, during which baryons are subject to complex physical processes such as heating, cooling, and mass inflow and outflow driven by AGN activity, supernova explosions, star formation, bulk motions \citep[e.g.,][]{Cen2006, Martizzi2019}, remain a mystery due to a lack of constraints from high-resolution spectroscopic observations.

It is expected that the advent of the next generation of X-ray missions, such as the X-Ray Imaging and Spectroscopy Mission (XRISM; \citealt{XRISMScienceTeam2020}), Hot Universe Baryon Surveyor (HUBS; \citealt{Cui2020}), Line Emission Mapper (LEM; \citealt{Kraft2022}), and Advanced Telescope for High ENergy Astrophysics (Athena; \citealt{Barret2020}), will provide powerful spectroscopic tools to achieve significant progress in understanding WHIM by making spectral diagnostics available with advance microcalorimetry technology. Among these, HUBS (expected to launch around 2030) is optimally designed for WHIM detection on scales of tens of megaparsecs thanks to its large effective area in the soft X-ray band, eV-level energy resolution, large field of view (FoV) of $1^\circ\times1^\circ$, and arcminute level angular resolution \citep{Cui2020}, which enables not only resolution of individual metal lines (e.g., O \textsc{viii}, O \textsc{vii}, Ne \textsc{ix}, Mg \textsc{xi}, Fe \textsc{xvii}, etc.) for unprecedented determination of WHIM properties, but also the line intensity mapping technique for enhanced detection of diffuse WHIM gas \citep{Bregman2023}, making direct detection of individual WHIM filaments feasible.

In this paper, we present our forecast study of the capability of HUBS in the detection and characterization of WHIM in cosmic filaments. We constructed a sample of inter-cluster filaments based on the first SRG/eROSTIA All-Sky Survey (eRASS1) supercluster catalog \citep{Liu2024} and modeled their emission properties (Sect. \ref{sec:sample}), and selected the most appropriate candidates for HUBS pointing observation (Sect. \ref{sec:select}). For these candidates, we created and analyzed mock spectra and images to forecast the scientific outcome to be achieved with HUBS (Sect. \ref{sec:mock}). Finally, we discuss the constraints on chemical origin, model assumptions, and existing optical observations in the fields of our candidate filaments (Sect. \ref{sec:discussions}), and we conclude our work (Sect. \ref{sec:concl}). Throughout this work, we adopt a flat $\Lambda$CDM cosmology with  $H_0$ = 67.4 km s$^{-1}$ Mpc$^{-1}$, $\Omega_\textrm{m}$ = 0.315, $\Omega_\Lambda$ = 0.6847, and $\Omega_\textrm{b}$  = 0.0493 \citep{PlanckCollaboration2020}. The ancillary response file (ARF) and redistribution matrix file (RMF) used for the mock observations are described in \cite{Zhang2022}, and were obtained from the HUBS team through private communications. We performed spectral modeling and fitting with \texttt{XSPEC v12.13} \citep{Arnaud1999} with the abundance table from \cite{Asplund2009} and atomic data from \texttt{AtomDB}\footnote{\url{https://atomdb.org/}} \texttt{v3.0.9}, and we quote 68\% confidence ranges unless otherwise stated.

\section{Inter-cluster filaments sample and emission model}\label{sec:sample}
\subsection{Sample construction}

Since the cosmic filaments are connected to galaxy clusters, superclusters (clusters of clusters) can provide hints for the soft X-ray emission of WHIM gas in filaments. Therefore, we built an inter-cluster filament sample based on the eRASS1 supercluster catalog \citep{Liu2024}, which is the largest-to-date catalog of X-ray selected superclusters, comprised of 1338 supercluster systems (encompassing 3948 member clusters) in the western Galactic hemisphere. 

Based on the member cluster properties given in \cite{Liu2024}, we analyzed each member cluster in each supercluster system to locate the inter-cluster filament connecting the cluster itself and its nearest neighbor, and constructed a sample of inter-cluster filaments. For the cluster under examination (cluster 1) and each other member (cluster 2) in the supercluster system, we first determined if their projected inter-cluster region is completely contaminated with cluster emission with the following procedure: (1) we calculated the projected angular distance $\delta_\textrm{proj}$ between cluster 1 and cluster 2 based on the coordinates of their X-ray centers; (2) we deduced the upper limit of their virial radii by applying the relation $R_\textrm{vir} \approx 10^{1/3} R_{500}$ \citep{Reiprich2013} to the 1$\sigma$ upper limit of $R_{500}$ given in \cite{Bulbul2024}; (3) if $\delta_\textrm{proj}$ is smaller than the sum of the upper limits of the two virial radii (in angular size), we excluded the pair in subsequent analysis. Second, we calculated the 3D physical distance $D_\textrm{phys}$ of all the remaining pairs. The member with the smallest $D_\textrm{phys}$ to the cluster under examination is considered the nearest neighbor. Third, we calculated the 3D comoving distance between the nearest neighbor pairs $D_\textrm{C}$, and removed the pairs with $D_\textrm{C} > 35\ h^{-1}\ \textrm{Mpc}$ based on \cite{Colberg2005} who found in an N-body simulation that the probability of having filaments between clusters reduces with their separation distance, and drops to 10\% when the comoving distance reaches $35\ h^{-1}\ \textrm{Mpc}$.  Finally, the remaining pairs of nearest neighbors were considered to host inter-cluster filaments between the clusters. As a result,  we constructed a sample of 1577 inter-cluster filaments. Distributions of redshift and inter-cluster distance are plotted in Fig. \ref{fig:filamentStats}.

\subsection{Filament emission model}\label{sec:filamentmodel}
\subsubsection{Filament geometry}\label{sec:geometry}
\begin{figure*}
    \centering
    \includegraphics[width=0.9\hsize]{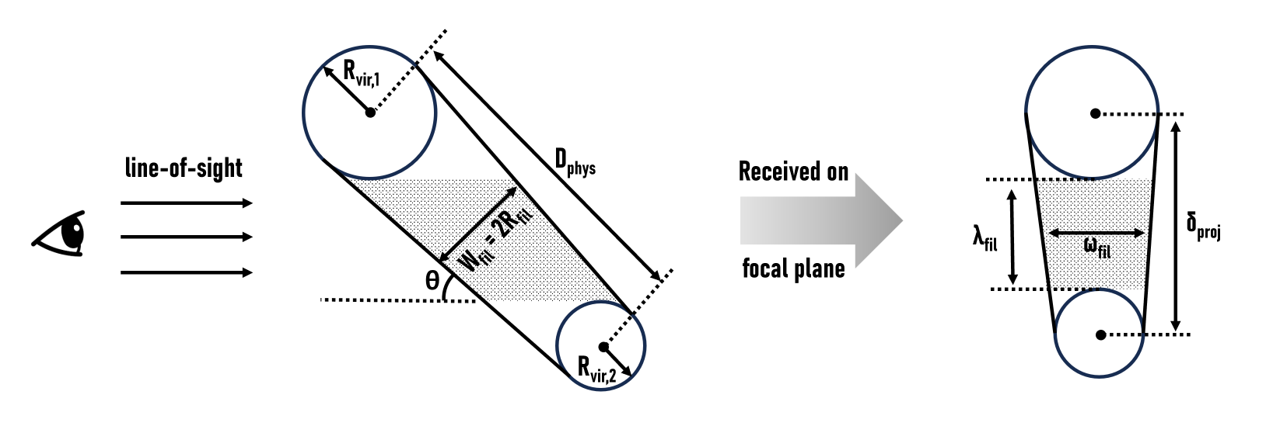}
    \caption{Geometry of modeled filaments. 
    $R_\textrm{vir,1}$ and $R_\textrm{vir,2}$ are virial radii of the clusters, $W_\textrm{fil}$ and $R_\textrm{fil}$ are filament width and radius respectively, $D_\textrm{phys}$ is the 3D physical distance between clusters, and $\theta$ is the inclination angle of the filament with respect to line-of-sight. After projected onto the focal plane, $\delta_\textrm{proj}$ is the separation between cluster centers, $\omega_\textrm{fil}$ is the filament width, and $\lambda_\textrm{fil}$ is the filament length after masking out virial regions of the clusters. The shaded areas represent the observed portion of the filament.}
    \label{fig:geometry}
\end{figure*}

We modeled the inter-cluster gas as cylinders directly connecting the two member clusters. It is worth noting that even though filaments bridging massive clusters were found to be approximately straight in simulations (e.g. \citealt{Cautun2014}), in reality the filament spine may deviate from being perfectly linear, so that the cylindrical assumption may cause reduced signal-to-noise ratio in stacking analysis. This has been discussed by \cite{Hoang2023}, who suggested that detection significance could be decreased due to the spatial alignment of filaments not following the axis of cluster pairs, after failing to detect excess emission in the stacked inter-cluster regions of 106 cluster pairs. Fortunately, it will not be a limitation for HUBS, as it is planned to perform pointing observations with a large FoV and fully cover each target filament individually so that a bent alignment will not cause degradation in the signal quality.

Fig. \ref{fig:geometry} shows a sketch of the model geometry of an inter-cluster filament connecting two clusters with virial radii $R_\textrm{vir,1}$ and $R_\textrm{vir,2}$, redshifts $z_1$ and $z_2$, and angular separation $\delta_\textrm{proj}$ between them.
For simplicity we assumed the cross-sectional radius of the filament cylinder to be the mean of the virial radii
\begin{equation}\label{eq:Rfil}
    R_\textrm{fil} = (R_\textrm{vir,1} + R_\textrm{vir,2}) / 2.
\end{equation} 
Thus the corresponding angular width and length of the filament after masking out the virial regions of the clusters are
\begin{equation}
    \omega_\textrm{fil} = \frac{2\times R_\textrm{fil}}{D_{A,z_\textrm{avg}}}
    \quad\text{and}\quad
    \lambda_\textrm{fil} = \delta_\textrm{proj} - \frac{R_\textrm{vir,1}}{D_{A,z1}} - \frac{R_\textrm{vir,2}}{D_{A,z2}},
\end{equation}
respectively, where $D_{A,z1}$, $D_{A,z2}$, and $D_{A,z_\textrm{avg}}$ are angular diameter distances at corresponding redshifts.
The inclination angle of the filament spine with respect to line-of-sight direction is determined by 
\begin{equation}\label{inc}
    \sin(\theta) = \frac{\delta_\textrm{proj} / D_{A,z_\textrm{avg}}}{D_\textrm{phys}}.
\end{equation}

\subsubsection{Gas emission}\label{sec:srcmodel}
For a straightforward estimation of the emission properties of the filaments in our sample, we assumed the WHIM to be in collisional ionization equilibrium (CIE), and modeled the filament emission as
\begin{equation}\label{eq:fil}
    \texttt{Model}_\textrm{fil} =\ \texttt{TBabs}\ \times\ \texttt{apec}_\textrm{WHIM},
\end{equation}
where \texttt{TBabs} represents the photo-absorption caused by neutral hydrogen in the Galactic ISM along the line-of-sight and $\texttt{apec}_\textrm{WHIM}$ represents the intrinsic emission of WHIM gas in the filament. The $\texttt{apec}$ component is described by four parameters: plasma temperature $kT$, metal abundance $Z$, redshift $z$, and normalization of emission measure $norm$, as explained below.

\paragraph{Hydrogen column density $N_\textrm{H}$:}
Column density of the foreground absorbing gas is set to be the total Galactic hydrogen column density at the central coordinates of each cluster pair. It was obtained with the \texttt{gdpyc}\footnote{\url{gdpyc.readthedocs.io}} tool following the \cite{Willingale2013} method, which considered both atomic hydrogen using the Leiden/Argentine/Bonn (LAB) 21-cm map \citep{Kalberla2005} and molecular hydrogen determined from the dust map of \cite{Schlegel1998}.

\paragraph{Temperature $kT$:}
Hydrodynamical simulations have revealed that WHIM gas temperature in the filaments may vary over a wide range (0.05 - 0.3 keV) across different codes and under different feedback scenarios. Meanwhile, a consensus on an isothermal core within a radius of $\sim1.5$ Mpc from the filament spine has been reached in various recent numerical works \citep{Gheller2019, Tuominen2021, Galarraga-Espinosa2021}. Therefore, a single-temperature model was adopted in our work.
Considering recent stacking results (e.g. 0.3 keV from \citealt{deGraaff2019}, 1.0 keV from \citealt{Tanimura2022}, and 0.5 keV from \citealt{Zhang2024}) and the measurement of individual filaments (e.g, 1.0 keV from \citealt{Veronica2024}), we set the $kT=0.5$ keV as the nominal case.

\paragraph{Metallicity $Z$:}
Though simulations \citep[e.g.,][]{Cen2006,Rahmati2016,Martizzi2019} have shown that the WHIM should have been efficiently polluted with metals, there have been no direct metallicity measurements for the filament gas. According to simulation results of \cite{Cen2006} (0.18 solar for oxygen in WHIM), \cite{Dave2007} ($\sim 0.2$ solar for shock-heated IGM at $z=0$), \cite{Biffi2022} ($\sim 0.2$ solar for the ``bridge'' between clusters), and observations of galaxy cluster outskirts (e.g., \citealt{Werner2013, Urban2017, Mantz2017, Biffi2018, Sarkar2022a}, $\sim 0.3$ solar for iron), we conservatively set $Z=0.2$ solar as the nominal case. 

\paragraph{Redshift $z$:}
Redshift of the filament was set to be the average of redshifts of the cluster pair as $z_\textrm{avg}$.

\paragraph{Normalization $norm$:}
The normalization parameter is defined as
\begin{equation}\label{eq:norm}
    norm = \frac{10^{-14}}{4\pi[D_A (1+z)]^2}\int{n_\textrm{e} n_\textrm{H} \textrm{d}V},
\end{equation}
where $D_A$ is the angular diameter distance to the source, $dV$ is the volume element, and $n_e$ and $n_\textrm{H}$ are electron and hydrogen densities, respectively. For the ionization-balanced plasma with our nominal temperature and abundance, we calculated the ratio of electron to hydrogen density to be $n_\textrm{e}/n_\textrm{H} \approx 1.172$.

The radial distribution (i.e., along the perpendicular direction to the filament spine) of gas overdensity is found to follow a $\beta$-like profile in simulations \citep{Gheller2019, Tuominen2021, Galarraga-Espinosa2021}, which is consistent with surface brightness profiles obtained from stacked observations \citep{Tanimura2022,Zhang2024}. We adopted the $\beta$-model for the radial distribution of overdensity $\delta$ \citep{Galarraga-Espinosa2022}:
\begin{equation}\label{eq:beta}
    1 + \delta(r) = \frac{1 + \delta_0}{{(1+{(\frac{r}{r_\textrm{c}})}^{\alpha})}^{\beta}},
\end{equation}
where $r$ is the radial distance from the filament spine, $r_\textrm{c}=0.56$ Mpc is the core radius, and $\alpha=1.57$ and $\beta=1.00$ are the slope indices for filaments connecting to massive clusters \citep{Galarraga-Espinosa2022}. We adopted a central overdensity $\delta_0=200$ for the nominal case \citep[e.g.,][]{Eckert2015,Veronica2024,Zhang2024, Dietl2024} and assumed a negligible evolution of overdensity in filaments in the redshift range \citep{Cui2019,Galarraga-Espinosa2024}. Then we calculated the electron density:
\begin{equation}
    n_\textrm{e}(r,z) = \frac{\delta(r)\rho_\textrm{crit}(z)\Omega_\textrm{b}(z)}{\mu_\textrm{e}m_\textrm{p}},
\end{equation}
where $\rho_\textrm{crit}(z)$ and $\Omega_\textrm{b}$ are the critical density and baryon fraction at redshift $z$, $\mu_\textrm{e} \approx$ 1.145 is the mean molecular weight per electron calculated for our nominal metallicity, and $m_\textrm{p} = 1.6726 \times 10^{-24}$ g is the proton mass. 

\begin{figure}
    \centering
    \includegraphics[width=0.9\linewidth]{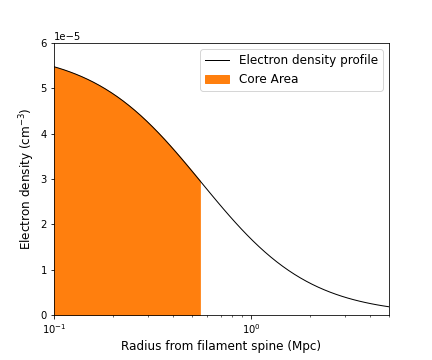}
    \caption{Gas density distribution in the direction perpendicular to the filament spine. Orange filled area indicates the core region ($r_c=0.56$ Mpc).}
    \label{fig:Core}
\end{figure}

With the centrally peaked density profile, the detected flux is higher towards the filament spine. For an optimal signal-to-background ratio, it's preferable to extract the spectra only from the core region of the filament. Therefore, we integrated the electron density over the core radius (0.56 Mpc, according to \citealt{Galarraga-Espinosa2022}) for each filament, as illustrated in Fig. \ref{fig:Core}, to calculate the $norm$ parameter. 

The velocity dispersion of the WHIM was ignored in our model due to the negligible spectral broadening expected for the energy band of interest. It was found in \cite{Rost2024} that the shock-heated filament gas has a velocity dispersion of $\sim500$ km/s, which corresponds to spectral broadening of $< 2$ eV at 1 keV.

\section{Candidate filaments}\label{sec:select}

From the 1577 filaments in our sample, we selected the candidate targets for HUBS pointing observations based on a series of criteria. The following criteria were set to ensure the significance of signal, confidence in the model, level of background emission, and geometrical suitability to HUBS specifications.

\subsection{Oxygen line significance}\label{sec:SNR}
With the eV-level energy resolution of HUBS, it is possible to effectively distinguish the emission of O \textsc{viii} Ly$\alpha$ line (0.654 keV) and O \textsc{vii} triplets (resonance, \textit{r}: 0.574 keV; inter-combination, \textit{i}: 0.569 keV; forbidden, \textit{f}: 0.561 keV) for sources with proper redshift so that these important emission lines are redshifted to energies relatively clean from the background emission lines \citep{Zhang2022}. The signal-to-noise ratio ($S/N$) of an emission line can be calculated from the source and background counts considering Poisson noise:
\begin{equation}\label{sigma}
    S/N= \frac{C_\textrm{src}}{\sqrt{C_\textrm{total}}} = \frac{R_\textrm{src} \times t_\textrm{exp}}{\sqrt{(R_\textrm{src} + R_\textrm{bkg}) \times t_\textrm{exp}}}, 
\end{equation}
where $C_\textrm{src}$ and $C_\textrm{total}$ are the source and total counts received, directly derived from exposure time $t_\textrm{exp}$ as well as source and background count rates ($R_\textrm{src}$ and $R_\textrm{bkg}$, respectively).

In the nominal case with gas temperature set at 0.5 keV, the O \textsc{viii} Ly$\alpha$ line has emissivity of about ten times of that of the O \textsc{vii} resonance line. Thus we chose the $S/N$ of O \textsc{viii} as the indicator for the detectability of each filament. Following the filament emission model in Sect. \ref{sec:srcmodel}, we obtained the O \textsc{viii} flux in $\textrm{photons}\ \textrm{arcmin}^{-2}\ \textrm{cm}^{-2}\ \textrm{s}^{-1}$ within a 6 eV wide band around the redshifted line energy, and multiplied it by the HUBS effective area in $\textrm{cm}^{2}$ at the corresponding energy to get the source count rate for each detector pixel (1 arcmin$^2$). To calculate the background count rates, we followed the prescription given in \cite{McCammon2002} to model the astrophysical background, which is comprised of the diffuse X-ray emission due to the cosmic X-ray background (CXB), the Milky Way Halo (MWH), and the Local Hot Bubble (LHB). A thorough description of our background model \texttt{Model}$_\textrm{bkg}$ and its parameters are given in Appendix \ref{sec:bkgmodel}.

The following criteria on exposure time, integration area, and $S/N$ were applied with regard to the O \textsc{viii} line:
\begin{itemize}
    \item A 200 ks exposure time is deemed reasonable for pointing observations, and therefore adopted.
    \item In order to achieve moderate spatial resolution with the O \textsc{viii} line, an integration area of 20 pixels (20 arcmin$^2$) is used.
    \item Assuming the above-mentioned exposure time and integration area, the filament should achieve $S/N>5$ for the O \textsc{viii} line.
\end{itemize}

For each of the 1577 filaments, we calculated the O \textsc{viii} $S/N$ with 200 ks exposure time 20 arcmin$^2$ integration area (Table \ref{tab:sampleproperties}). After applying the $S/N$ criterion, 283 filaments remained.

\subsection{Confidence in model parameters}
Our calculation in Sect. \ref{sec:SNR} is heavily dependent on the model described in Sect. \ref{sec:filamentmodel}. While many aspects of model uncertainties apply to all filaments in the sample, calculation of the emission normalization parameter is strongly dependent on the clusters' redshift. Furthermore, the assumption of filament temperature can be related to cluster mass. Therefore, we filtered out filaments whose connected clusters have large redshift uncertainties or small cluster masses. 

According to Eq. \ref{eq:norm}, the calculation of emission normalization is critically determined by filament length $D_\textrm{phys}$ and the line-of-sight inclination angle $\theta$, both of which rely on the redshift measurement of the connected clusters. Based on \cite{Kluge2024}, we identified the galaxy clusters with spectroscopic redshift measurement (determined by either the redshifts of at least three spectroscopic members or the spectroscopic redshift of the optical center). To evaluate the error in emission normalization, we simulated each filament 10000 times with the redshift error of each cluster in the pair. We find that for filaments with both connected clusters having spectroscopic measurements, the median error in emission normalization is 14.3\%, and for filaments with only one or no connected clusters having spectroscopic measurements, the median error in emission normalization is 53.4\% and 59.8\% respectively. For this reason, we selected filaments only when both of its connected clusters are labeled with spectroscopic redshift. After applying this criterion, 41 filaments remained. 

The density and temperature of our filament model are based on results of recent observations, but some simulation works such as \cite{Angelinelli2021} have found that filament temperature can be positively correlated with the mass of the cluster connected to it. Furthermore, \cite{Galarraga-Espinosa2020} found in their analysis on the Illustris-TNG simulations that filaments connected to maximum critical points with $M_{200} > 10^{14.05} M_\odot$ belong to the hot and dense filament population. Therefore, we selected filaments only when both clusters in the pair have $M_{500} > 10^{14} M_\odot$, which is a common threshold adopted in many simulation works that analyzed filaments connected to galaxy clusters \citep[e.g.][]{Gonzalez2010, Aragon-Calvo2010,Pereyra2020}. After applying this criterion, 38 filaments remained.

\subsection{Background emission}
To avoid regions of high background levels, we checked the central coordinates of the filaments to exclude those located on the eROSITA bubbles, which are the soft X-ray excess extending to about 14 kpc above and below the Galactic plane \citep{Predehl2020}. For each of the remaining filaments, we also checked the X-ray background count rate measured in the R45 (3/4 keV) band of ROentgen SATellite (ROSAT) All-Sky Survey \citep{Snowden1997} using the \texttt{sxrbg} tool \citep{Sabol2019}. We find that for all of the filaments located on the eROSITA bubble, background count rates averaged within the cone region (1 degree radius of the filament central coordinates) exceed $150 \times 10^{-6}$ counts s$^{-1}$ arcmin$^{-2}$, while those not on the eROSTIA bubble have background count rates below this threshold. 
After applying this criterion, 28 filaments remained.

\subsection{Consideration on contamination sources}
To avoid line-of-sight contamination of diffuse X-ray emitting sources, we inspected the filament coordinates to make sure there were no foreground or background clusters directly between the connecting clusters or in the vicinity of the expected filament spine. On the other hand, clumps within the filaments can also cause severe contamination in spectral analysis, as the clump gas is hotter and denser than the filament gas. In the stacked filament analysis of \cite{Zhang2024}, galactic halos with the central galaxy's stellar mass within $10^{11} M_\odot < M_\star < 10^{12} M_\odot$ were found to contribute $\sim20\%$ of soft X-ray emission at filament locations. 
In order to allow for effective identification and removal of these clumps in practice, we require the angular size of the filament to be sufficiently large so that angular variation of the soft X-ray emission can be analyzed. Because the HUBS angular resolution is 1 arcmin, we require the projected core width ($\omega_\textrm{core}$, 0.56 Mpc) to be larger than 5 arcmins and the projected filament length ($\lambda_\textrm{fil}$) to be larger than 20 arcmins to ensure adequate area when masking is required.

\subsection{Selected candidates}
Finally, we selected four candidate filaments that met all of our criteria, with their connecting clusters and filament properties listed in Tables \ref{tab:clusterproperties} and \ref{tab:filamentproperties}, respectively. Since the filaments were selected from supercluster systems and no two candidates fall in the same supercluster, in this paper we adopt the names of the corresponding superclusters as the unique identifiers for the candidate filaments. Meanwhile, relevant properties of all 1577 filaments in the sample can be found in Table \ref{tab:sampleproperties}.

Among the candidate filaments, the highest O \textsc{viii} $S/N$ (6.03) is expected from J1214+2712, which also processes the largest 3D comoving distance (28.3 $h^{-1}\ \textrm{Mpc}$ and largest 3D physical distance (37.8 Mpc) between clusters, the largest projected length (31.33 arcmin), and smallest inclination angle (14.6 degrees). Three out of the four candidates are in the northern hemisphere, with only J2300-4534 in the southern hemisphere. 

\begin{table*}
\caption{Properties of connecting clusters for the candidate filaments.}
\label{tab:clusterproperties}
    \centering
    \begin{tabular}{c|cccccc}
        \hline 
        \hline

        \multirow{2}{*}{Filament Name}&\multirow{2}{*}{Cluster Name}& $z$&$M_{500}$&  $R_{500}$&$R_\textrm{vir}$  &$kT$\\
        & & (-)& ($10^{14} M_\odot$)&  (Mpc)&(Mpc) &(keV)\\
        \hline
        \multirow{2}{*}{J0935+1949}&1eRASS J093552.6+201132
&$0.1757\pm0.0004$&$1.54_{-0.38}^{+0.32}$& $0.77_{-0.07}^{+0.05}$
&$1.66_{-0.15}^{+0.11}$ &$1.66_{-0.53}^{+2.24}$\\
        &1eRASS J093446.5+192751
&$0.1807\pm0.0004$&$1.17_{-0.44}^{+0.42}$&  $0.7_{-0.1}^{+0.08}$
&$1.51_{-0.22}^{+0.16}$ &$>1.08$\\
        \hline
        \multirow{2}{*}{J0945+3418}&1eRASS J094554.6+344046
&$0.1344\pm0.0004$&$1.41_{-0.31}^{+0.32}$& $0.76_{-0.06}^{+0.05}$
&$1.63_{-0.13}^{+0.12}$ &$>1.52$\\
        &1eRASS J094426.2+335718
&$0.1306\pm0.0004$&$1.91_{-0.34}^{+0.4}$&  $0.84_{-0.05}^{+0.05}$
&$1.81_{-0.12}^{+0.12}$ &$1.51_{-0.34}^{+1.49}$\\
        \hline
        \multirow{2}{*}{J1214+2712}&1eRASS J121411.6+263822
&$0.1733\pm0.0038$&$1.36_{-0.29}^{+0.26}$& $0.74_{-0.06}^{+0.04}$
&$1.59_{-0.12}^{+0.10}$&$>1.28$\\
        &1eRASS J121259.1+272708
&$0.1797\pm0.0003$&$3.36_{-0.29}^{+0.44}$&  $1.00_{-0.03}^{+0.04}$&$2.15_{-0.06}^{+0.09}$ &$1.23_{-0.16}^{+0.45}$\\
\hline
 \multirow{2}{*}{J2300-4534}& 1eRASS J230541.6-451201
& $0.1275\pm0.0008$
& $4.15_{-0.43}^{+0.32}$
&  $1.09_{-0.04}^{+0.03}$
&$2.35_{-0.09}^{+0.06}$
 &N/A\tablefootmark{*}\\
 & 1eRASS J230557.7-442221
& $0.1318\pm0.0011$
& $2.00_{-0.27}^{+0.34}$&  $0.85_{-0.04}^{+0.05}$
&$1.84_{-0.08}^{+0.1}$
 &$0.38_{-0.05}^{+0.21}$\\
\hline
    \end{tabular}
\tablefoot{\\Redshift $z$, mass $M_{500}$, radius $R_{500}$ and cluster temperature $kT$ and corresponding 1$\sigma$ errors were obtained from \cite{Bulbul2024}, while $R_\textrm{vir}$ is derived based on $R_\textrm{vir} \approx 10^{1/3} R_{500}$ given in \cite{Reiprich2013}.
\\
\tablefootmark{*}Temperature of 1eRASS J230541.6-451201 provided in \cite{Bulbul2024} was unrealistic.  }
\end{table*}

\begin{table*}
    \caption{Physical and emission properties of candidate filaments.}
    \label{tab:filamentproperties}
    \centering
    \begin{tabular}{c|cc|ccccccc|c}
    \hline\hline
         \multirow{2}{*}{Filament Name}&   $z_\textrm{avg}$&$N_\textrm{H}$& $D_\textrm{C}$&$D_\textrm{phys}$ & $\delta_\textrm{proj}$&$\omega_\textrm{fil}$&$\omega_\textrm{core}$&   $\lambda_\textrm{fil}$&$\theta$&O \textsc{viii} $S/N$\\
         &   (-)&($10^{20}$ cm$^{-2}$)& ($h^{-1}$ Mpc)&(Mpc) & (deg)&(arcmin)&(arcmin)&   (arcmin)&(deg)&(-)\\
         \hline
         J0935+1949
&   0.1782
&3.68& 22.7&30.2
& 0.77
&16.91
&5.98
&    29.45
&16.7
&5.15
\\
         J0945+3418
&   0.1325
&1.15& 17.6&21.9
& 0.78
&23.49
&7.64
& 
 
  23.49
&18.3
&5.12
\\
         J1214+2712
&   0.1765
&1.85& 28.3&37.8
& 0.86
&20.12
&6.03
&    31.33
&14.6
&6.03
\\
 J2300-4534& 0.12965
& 1.06
&  19.7&24.4
& 0.83
& 29.07
& 7.79
& 20.97
& 17.1
&5.56
\\
\hline
    \end{tabular}
\tablefoot{
$z_\textrm{avg}$ and $N_\textrm{H}$ were obtained following Sect. \ref{sec:srcmodel}, $D_\textrm{C}$ is the comoving distance, $D_\textrm{phys}$, $\delta_\textrm{proj}$, $\omega_\textrm{fil}$, $\lambda_\textrm{fil}$, and $\theta$ were described in Fig. \ref{fig:geometry}, $\omega_\textrm{core}$ is the angular size corresponding to the core radius $r_\textrm{c}=0.56$ Mpc, and O \textsc{viii} $S/N$ was calculated based on Sect. \ref{sec:SNR}.
}
\end{table*}

\section{Simulation and analysis of mock observations}\label{sec:mock}
\subsection{Spectral diagnostic capabilities}\label{sec:specmock}
\subsubsection{Simulation and fitting procedure}
In order to quantify the level of diagnosis of the gas properties to be achieved by HUBS, we generated and analyzed mock spectra for the candidate selected in Sect. \ref{sec:select}.
Assuming 200 ks exposure time for each of the candidate filaments, we used the \texttt{XSPEC} tool \texttt{fakeit} to simulate the spectra including both \texttt{Model}$_\textrm{fil}$ and \texttt{Model}$_\textrm{bkg}$ as described in Sect. \ref{sec:srcmodel} and Appendix \ref{sec:bkgmodel} respectively.
Poisson errors were assumed for statistical uncertainties, while the consideration of systematic uncertainties from instrumental calibration (i.e., effective area, energy scale, and line spread function) is deferred to future works after the completion of technology development.

\begin{figure}
    \centering
    \includegraphics[width=0.9\linewidth]{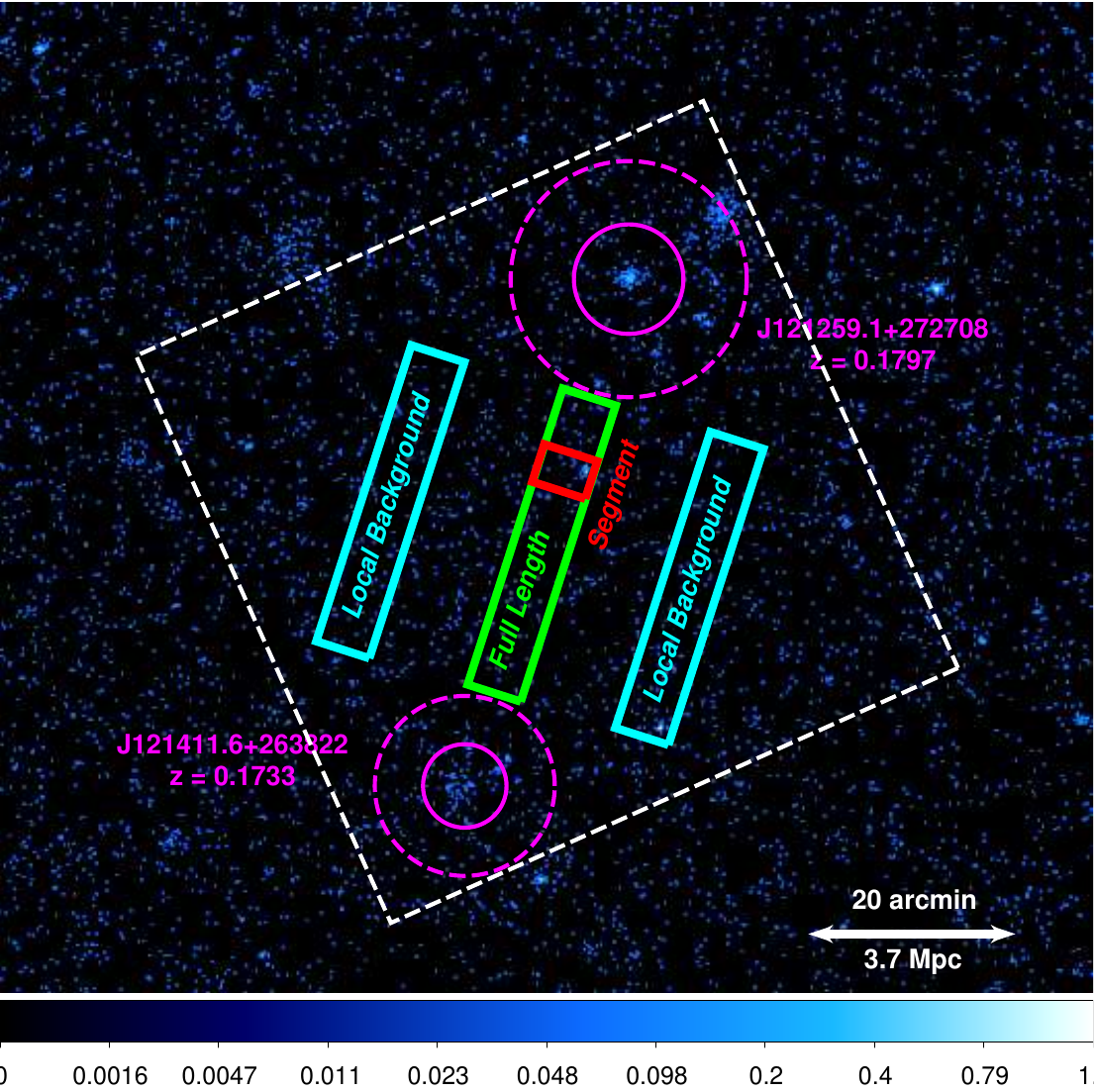}
    
    \caption{Regions used to extract filament and background spectra for the case of J1214+2712. The image was generated from $0.2-2.0$ keV event lists from eRASS1 data. The green solid box represents the ``full length'' region, the red solid box represents the ``segment'' region, and the cyan solid boxes represent the region used for ``local background'' extraction. The magenta solid and dashed circles represent R$_{500}$ and R$_\textrm{vir}$ of the clusters, and the white dashed box is HUBS FoV of $1^\circ\times1^\circ$.}
    \label{fig:extraction}
\end{figure}

First of all, the source spectra were extracted from a region covering the entire core area of $\omega_\textrm{core} \times \lambda_\textrm{proj}$ (denoted as ``full length'').
Secondly, though extremely difficult for diffuse emission of filaments, characterizing the spatial variation of gas properties along the filament spine can be beneficial for understanding the evolution of the large-scale structure and the cluster accretion process. Thus another extraction region covering a 20 arcmin$^2$ segment in the core (denoted as ``segment'') was also simulated to test the capability of diagnosing spectral properties with moderate spatial resolution. Our candidate filaments have core areas of $160-190$ arcmin$^2$, so that in practical observations about eight to ten ``segment'' regions can be extracted from each filament.
Finally, a ``local background'' spectrum was also simulated for each filament using the \texttt{Model}$_\textrm{bkg}$ only, with the extraction region being twice the area of that of ``full length''. The extraction regions of source and background spectra are indicated in Fig. \ref{fig:extraction}.

The simulated spectra were grouped to a reduced number of energy bins following \cite{Kaastra2016} using the \texttt{ftgrouppha} tool. Fitting was applied to the simulated data within the rest-frame energy range of $0.5-1.4$ keV, where source emission is not overwhelmed by the background and can be easily separated. 
We assume the local background can be well established in the pointing observation, so we fitted the data with the \texttt{TBabs} $\times$ (\texttt{apec} + \texttt{powerlaw}) model, where the \texttt{apec} component represents the WHIM gas emission and the \texttt{powerlaw} component accounts for any residuals in the CXB. 
The temperature, metallicity, and normalization of \texttt{apec}, as well as normalization of \texttt{powerlaw}, were allowed to vary freely, while other parameters (absorption column density, redshift, and powerlaw slope) were fixed at model values. 
We applied C-statistics \citep{Cash1979} instead of ${\chi}^2$-statistics \citep{Lampton1976} to obtain unbiased estimates of model parameters for data with a low count number in each bin \citep{Kaastra2017}. After an initial fitting, we ran 
a Monte Carlo Markov Chain (MCMC) using the initial fit results of the free parameters as priors to determine the final best-fit values and uncertainty ranges. 

\begin{figure}
    \centering
    \includegraphics[width=1.0\linewidth]{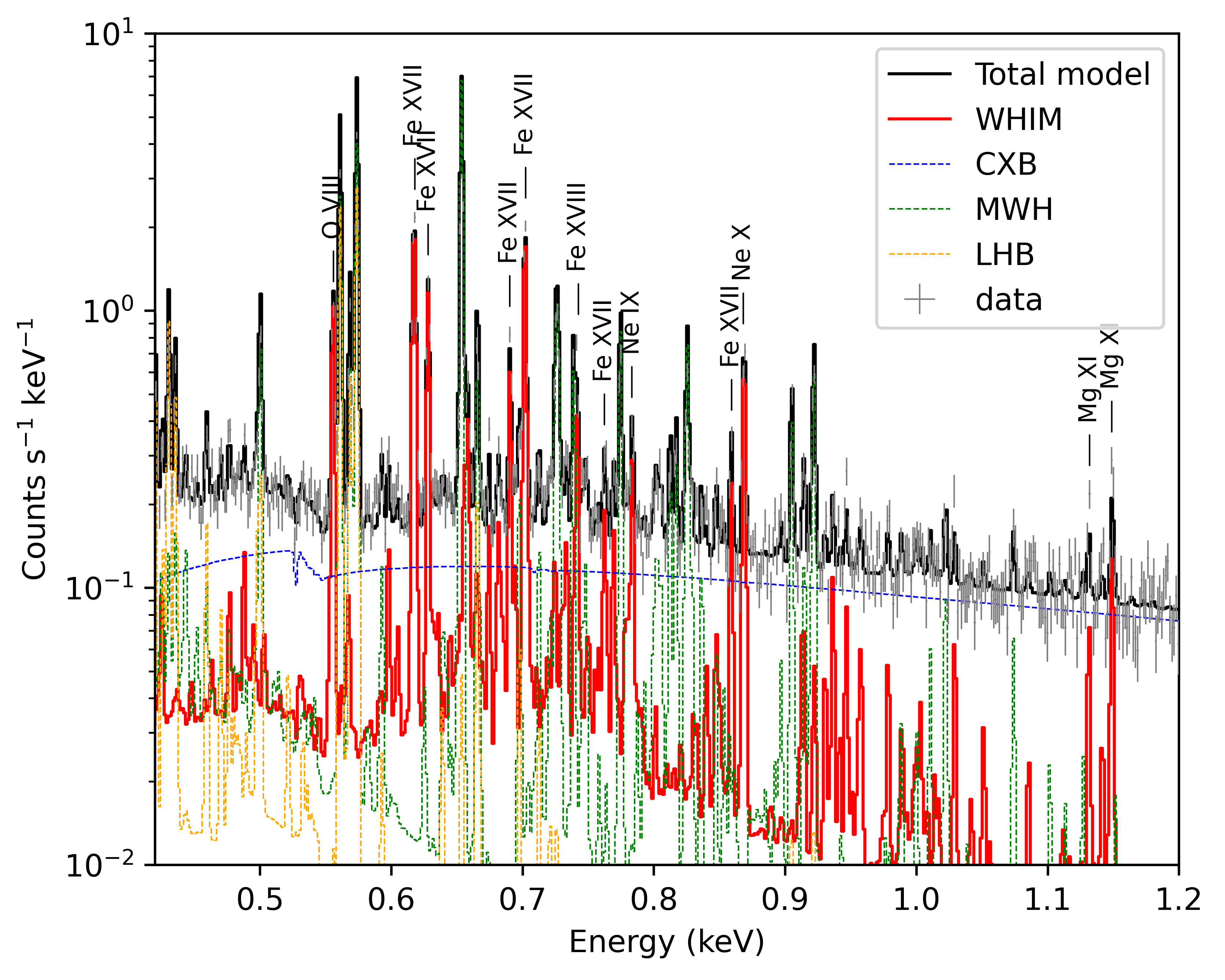}\\
    \vspace{0.2cm}
    \includegraphics[width=1.0\linewidth]{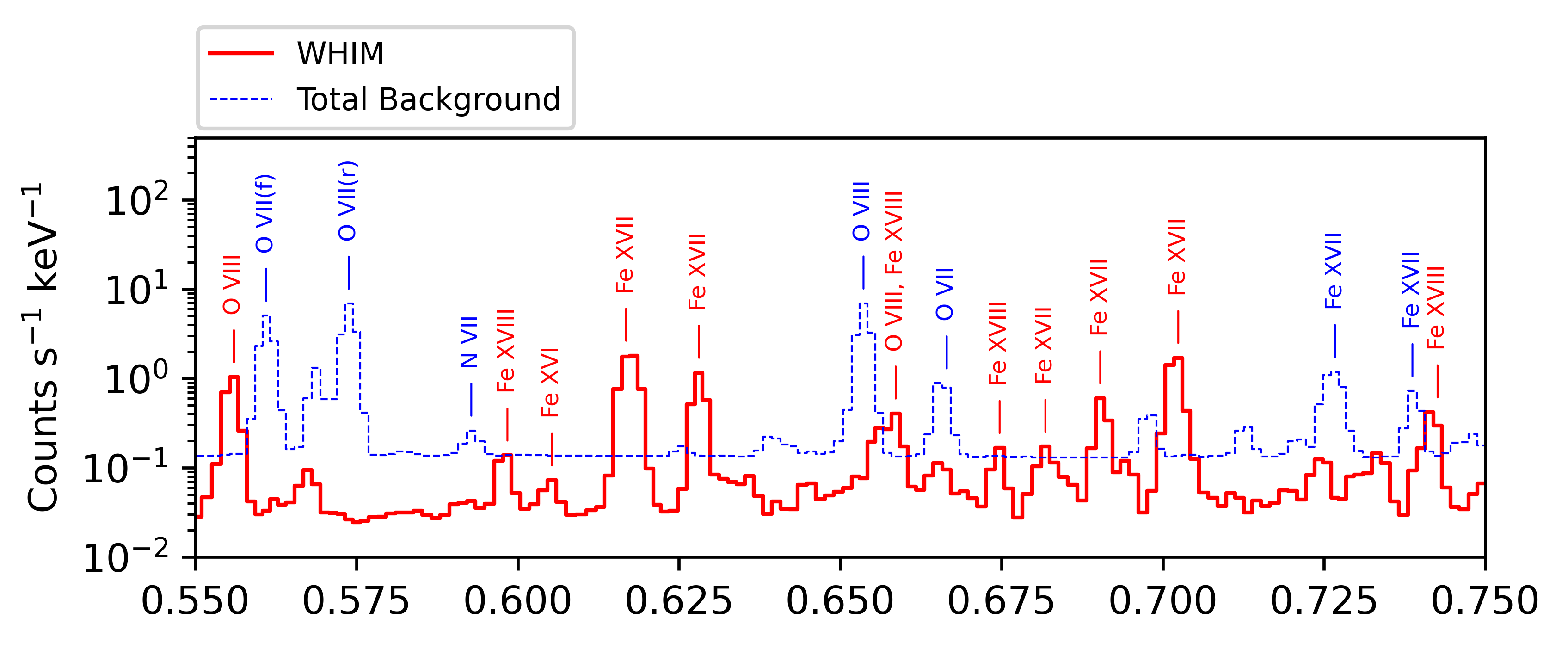}
    \includegraphics[width=1.0\linewidth]{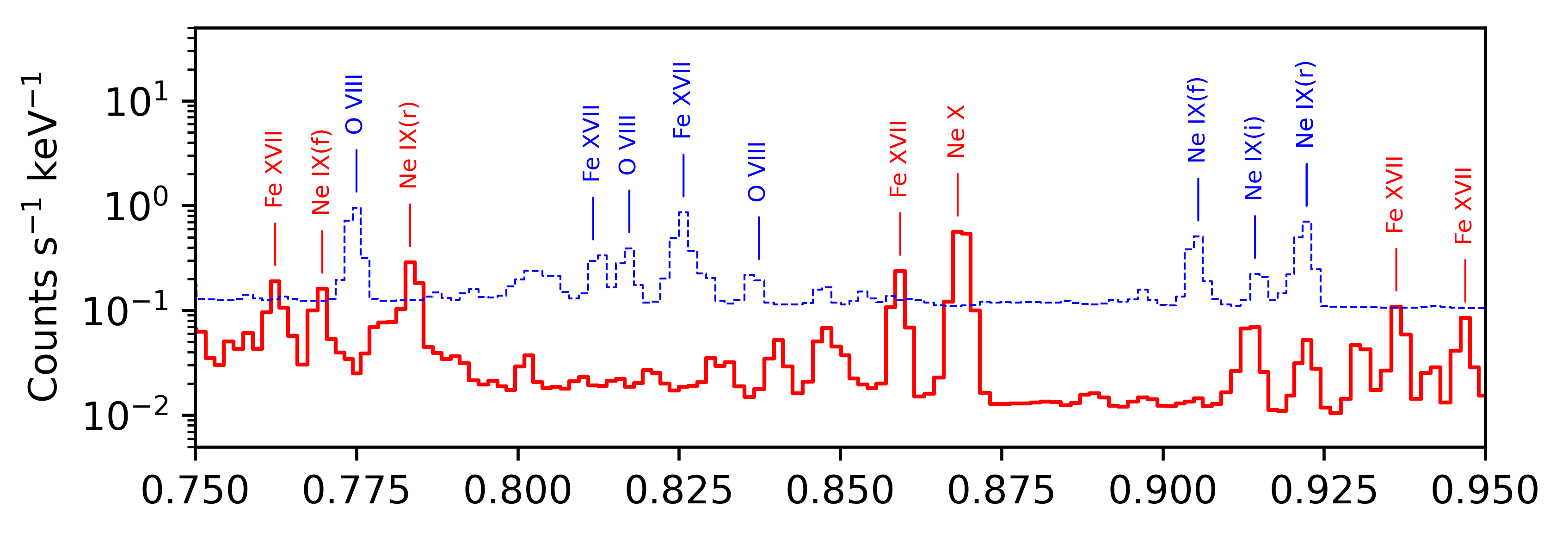}
    \includegraphics[width=1.0\linewidth]{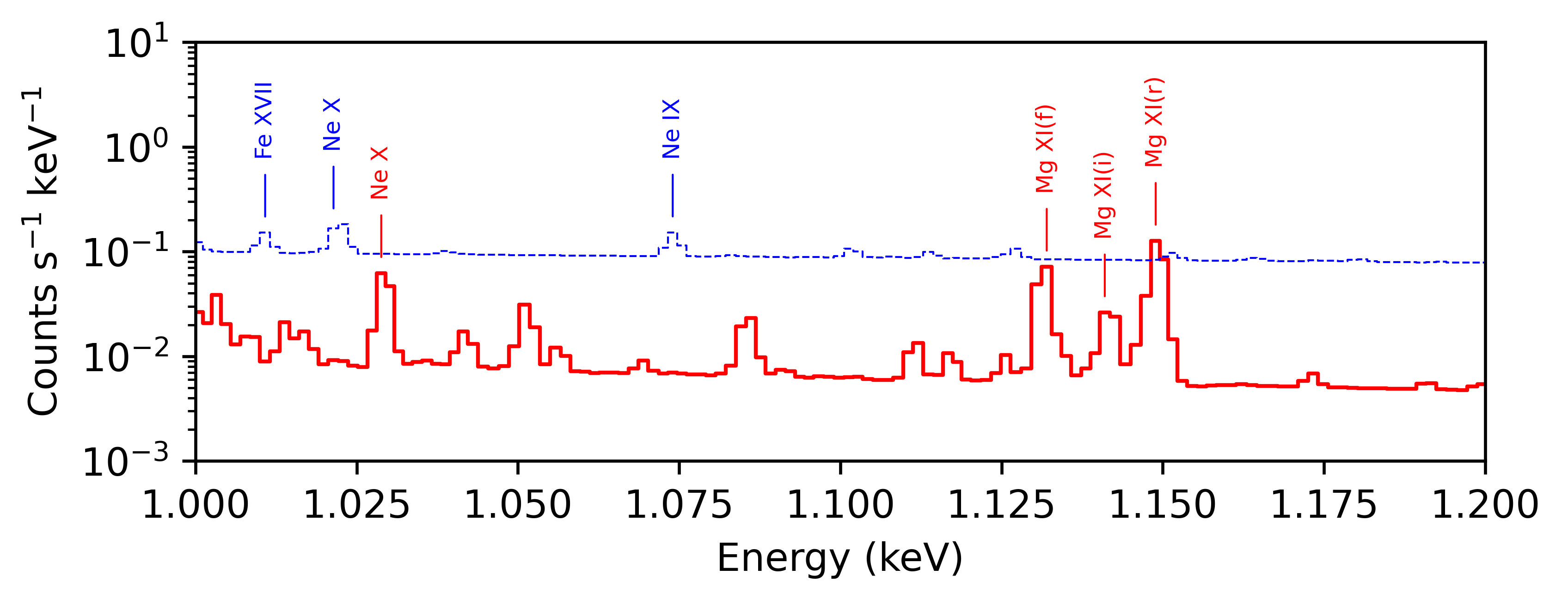}
    \caption{Mock spectra of the ``full length'' region of J1214+2712. \textit{Top panel:} Mock data (grey crosses) and best-fit model (black solid line). Components of the model are also shown (WHIM in red solid line, CXB in blue dashed line, MWH in cyan dashed line, and LHB in orange dashed line). \textit{Bottom panel:} Zoom-in spectra of the WHIM (red solid line) and the total background (blue dashed line) with prominent emission lines labeled.}
    \label{fig:mockspectrafit}
\end{figure}

An example of the mock spectra and model components is shown in Fig. \ref{fig:mockspectrafit} for J1214+2712. Thanks to the small inclination angle $\theta$ of each candidate, emission of the diffuse WHIM can be integrated over a large volume for a given angular area, thus the observed intensities of certain redshifted metal lines including O \textsc{viii}, Fe \textsc{xvii}, Fe \textsc{xviii}, Ne \textsc{ix}, Ne \textsc{x}, and Mg \textsc{xi} can exceed the background emission, allowing for spectral diagnostics of gas properties. 

\subsubsection{Fitting results}
\begin{table*}
    \caption{Best-fit results for the HUBS mock spectra under the nominal case using the \texttt{apec} model. The mock exposure time is 200 ks.}
    \label{tab:apec_fit}
    \centering
    
    \begin{tabular}{ccccccll}
        \hline\hline
         \multirow{2}{*}{Filament Name}&  $kT$\tablefootmark{a}&  $Z$\tablefootmark{a}&  $norm$\tablefootmark{a}&   C-stat/dof  &Net counts\tablefootmark{b}& $\delta_\textrm{offset}$ \tablefootmark{c}& $\delta_\textrm{error}$\tablefootmark{c}\\
 & (keV)& (solar)& ($10^{-5}$ cm$^{-5}$)&   &  & & \\
 \hline
 \multicolumn{8}{c}{``full length''}\\
 \hline
         J0935+1949
&  $0.50_{-0.01}^{+0.01}$(0.5)&  $0.21_{-0.01}^{+0.03}$(0.2)&  $38.5_{-4.8}^{+0.9}$(39.3)&   522.7/524&8408& 1.0\%& $_{-6.5\%}^{+1.1\%}$
\\
         J0945+3418
&  $0.50_{-0.01}^{+0.01}$(0.5)&  $0.20_{-0.01}^{+0.02}$(0.2)&  $31.7_{-3.2}^{+1.6}$(30.9)&   549.5/545&8118& 1.4\%& $_{-5.2\%}^{+2.5\%}$
\\
         J1214+2712
&  $0.48_{-0.01}^{+0.01}$(0.5)&  $0.20_{-0.01}^{+0.03}$(0.2)&  $48.1_{-6.2}^{+2.6}$(49.1)&   455.9/531&11280& 1.0\%& $_{-6.7\%}^{+2.7\%}$
\\
         J2300-4534
&  $0.49_{-0.01}^{+0.01}$(0.5)&  $0.22_{-0.02}^{+0.03}$(0.2)&  $29.4_{-3.4}^{+3.4}$(31.7)&   554.4/542&8611& 3.7\%& $_{-6.0\%}^{+5.7\%}$
\\
\hline
         \multicolumn{8}{c}{``segment''}\\
         \hline
         J0935+1949
&  $0.51_{-0.03}^{+0.01}$(0.5)&  $0.16_{-0.02}^{+0.07}$(0.2)&  $5.0_{-1.4}^{+0.4}$(4.5)&   403.3/412&947& 5.5\%& $_{-15.3\%}^{+4.1\%}$
\\
 J0945+3418
& $0.48_{-0.02}^{+0.02}$(0.5)& $0.20_{-0.01}^{+0.20}$(0.2)& $3.5_{-1.7}^{+0.2}$(3.4)& 449.2/432& 953& 1.1\%& $_{-28.6\%}^{+3.1\%}$
\\
 J1214+2712
& $0.52_{-0.02}^{+0.02}$(0.5)& $0.18_{-0.00}^{+0.11}$(0.2)& $5.9_{-2.2}^{+0.1}$(5.2)& 375.7/418& 1320& 6.1\%& $_{-21.5\%}^{+1.3\%}$\\
 J2300-4534
& $0.50_{-0.02}^{+0.02}$(0.5)& $0.24_{-0.05}^{+0.22}$(0.2)& $3.2_{-1.5}^{+0.8}$(3.9)& 382.8/431& 1044& 8.8\%& $_{-27.3\%}^{+12.2\%}$
\\
 \hline
    \end{tabular}
\tablefoot{
Top panel shows the results of spectra extracted from the ``full length'' region and bottom panel shows results from the ``segment'' region.
\\
\tablefoottext{a}{Parameters $kT$, $Z$, and $norm$ given in best-fit value with 68\% confidence range, followed by the model value used for spectral simulation in parenthesis.}\\
\tablefoottext{b}{Net counts of the spectra in the $0.5-1.4$ keV rest-frame energy range after background subtraction.}\\
\tablefoottext{c}{$\delta_\textrm{offset}$ and $\delta_\textrm{error}$ are the deviation from simulation input value and 68\% error of the best-fit value of the gas density inferred from $norm$ fitting results following Eq. (\ref{eq:norm}) respectively.}
}
\end{table*}

Fitting results and 68\% confidence ranges of the \texttt{apec} parameters on the ``full length'' and ``segment'' spectra are listed in the top panel and bottom panel of Table \ref{tab:apec_fit} respectively. 
With normalization converted to gas density using Eq. (\ref{eq:norm}), we report two additional quantities: gas density offset ($\delta_\textrm{offset}$), which is the difference between the fitted and input density divided by the model density, and gas density error ($\delta_\textrm{error}$), which is the 68\% uncertainty for the fitted density.

We find the following for the 200 ks HUBS mock spectra simulated for the ``full length'' region:
\begin{itemize}
    \item Good fitting statistics were achieved for all candidates, indicated by the ratio of C-statistic value to degrees of freedom (C-stat/dof) $\sim1$.
    \item The best-fit results of all three \texttt{apec} parameters ($kT$, $Z$, and $norm$) are consistent with the corresponding input parameter in the simulation within 68\% confidence range for all filaments, indicating the 200 ks exposure time is sufficient to recover the physical properties of WHIM in filaments.
    \item 68\% error for metallicity are within $\pm0.03$ solar for all filaments, which will provide unprecedented constraints on the chemical status of WHIM gas.
    \item The derived constraints for gas density result in offsets of less than 5\% and \textrm{68\%} errors of less than $\pm10\%$ with respect to the input gas density for all candidates. 
\end{itemize}

Among all candidates, the highest net counts and best-fitting statistics were obtained with J1214+2712, which reinforces its status as the leading candidate. Specifically, 200 ks HUBS observation can constrain its metallicity with 68\% upper and lower errors of 0.03 and 0.01 solar, and gas density with 68\% upper and lower errors of +2.7\% and -6.7\% for J1214+2712.

\begin{figure}
    \centering
    \includegraphics[width=1.0\linewidth]{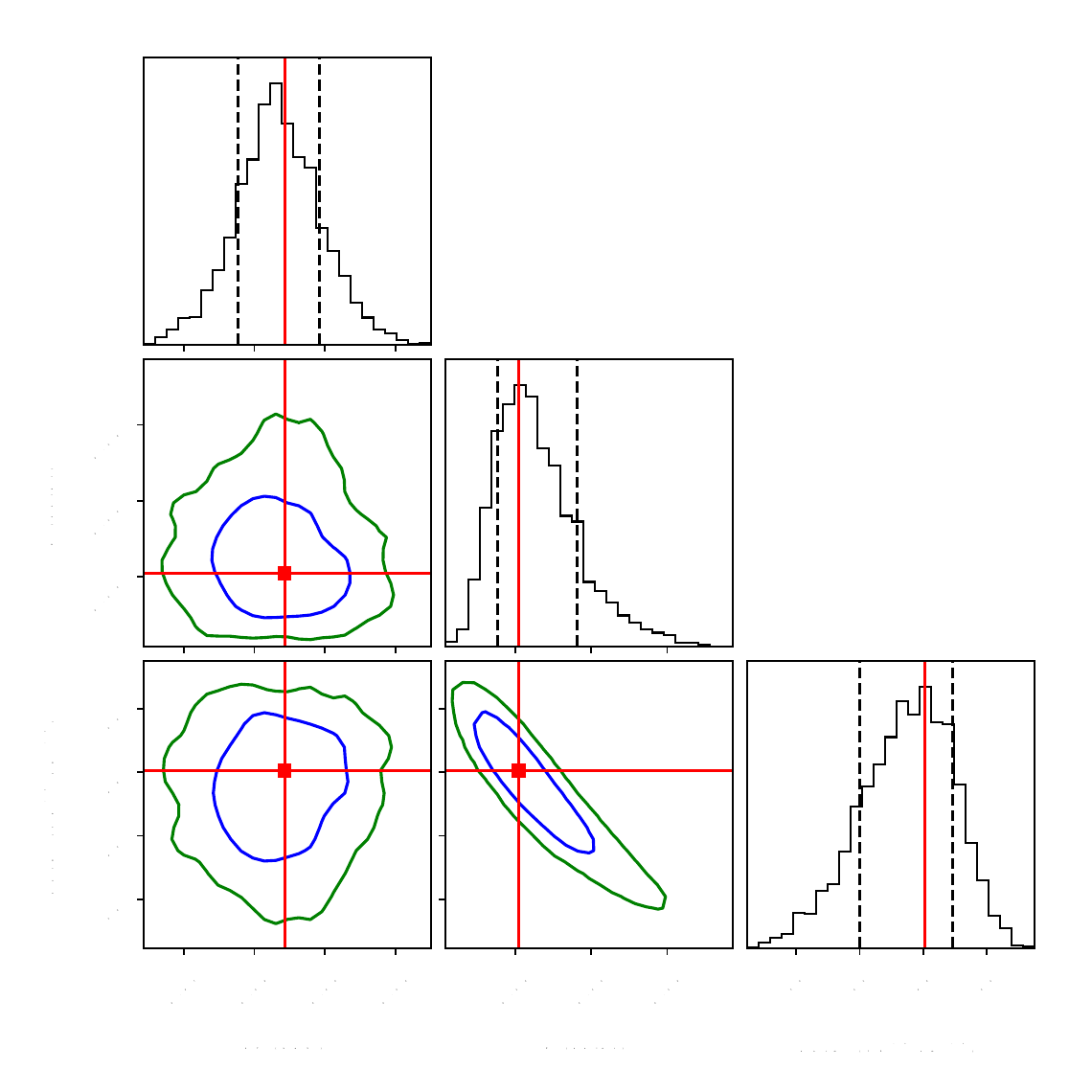}
    \caption{Probability density distributions of parameters $kT$, $Z$, and $norm$ of the ``full length'' 200 ks mock spectra of J1214+2712. The red points and lines represent the best-fit values. The blue and green contours represent 68\% and 95\% confidence levels. The dashed vertical lines on the 1-D histograms represent 68\% confidence interval of each parameter.}
    \label{fig:contour}
\end{figure}

We also find that the metallicty-density degeneracy was not severe for the mock spectra. The fraction of metal line emission over bremsstrahlung emission increases as the temperature of X-ray emitting plasma decreases (see Figure 18 of \citealt{Sutherland1993}), introducing a degeneracy between the metallicity and emission measure, which causes the measurement of gas density to experience large uncertainty when metallicity cannot be accurately determined \citep{Mernier2022, Gastaldello2021}. This is especially a problem for CCD-based spectrometers, whose $\sim100$ eV energy resolution cannot separate individual lines in the Fe-L complex as well as those source emission lines close to the background lines. The situation can be significantly improved with microcalorimeter-based spectrometers with eV-level energy resolution.
In the case of HUBS, we show the probability density distributions of the WHIM parameters in Fig. \ref{fig:contour}, which indicates that even with the metallicity-density degeneracy for WHIM temperature, good constraints for these parameters can still be obtained.

The ability to perform spatially resolved measurements of WHIM gas properties with HUBS on the candidate filaments can be implied from fitting results of ``segment'' spectra listed in the bottom panel of Table \ref{tab:apec_fit}. Even With reduced number of counts, the best-fit parameters are still consistent with input values within 68\% confidence range for all candidates. While the uncertainties of normalization and metallicity are significantly worse compared with the ``full length'' cases due to degeneracy (especially the lower bounds of $norm$ and the upper bounds of $Z$), precise constraints of temperature (68\% error of $\sim\pm0.02$ keV) can still be obtained. Despite the degradation in fitting results, the constraints can be improved to a similar level of ``full length'' cases by increasing exposure time by eight to ten times to recover similar net counts.

\subsection{Narrowband imaging capabilities}\label{sec:SOXSmodel}
In practical observations of inter-cluster filaments, it is crucial to trace the alignment of each filament in order to identify appropriate regions for spectral data extraction. Therefore, it is beneficial to directly map WHIM distribution for the candidate filaments so that the shape of each filament, which may deviate from a linear alignment, can be properly determined.
The narrowband imaging capability of HUBS has been demonstrated in \cite{Zhang2022} by simulating oxygen line images of the warm-hot ionized gas around distant galaxies, galaxy groups, and galaxy clusters with 1 Ms of deep exposure. Though even more diffuse, our candidate filaments are advantageous for their large physical lengths and small inclination angles along the line-of-sight which result in large emission measure normalized to angular size. 
We generated the narrow band O \textsc{viii} map of J1214+2712 to provide an example.

\subsubsection{Simulation procedure}

The mock observations were created by using \texttt{pyXSIM}\footnote{\url{https://hea-www.cfa.harvard.edu/~jzuhone/pyxsim/index.html}} \texttt{v4.4.0} and \texttt{SOXS}\footnote{\url{https://hea-www.cfa.harvard.edu/soxs/index.html}} \texttt{v4.8.5}. \texttt{pyXSIM} was used to generate the projected distribution of photons for our model and \texttt{SOXS} was used simulate the received events from the source photons using specified instrument files and background models.

The J1214+2712 filament, our most promising candidate, resides in the 1eRASS-SC J1214+2712 supercluster, connecting member galaxy clusters 1eRASS J121411.6+263822 ($z=0.1733$) and 1eRASS J121259.1+272708 ($z=0.1797$). The 1eRASS-SC J1214+2712 is a three-cluster system, with the third member being 1eRASS J21529.6+273142 ($z=0.1835$). Another cluster pair in the system comprised of J121259.1+272708 and J21529.6+273142 is also identified as an inter-cluster filament in our sample, and is expected to have O \textsc{viii} $S/N$ of 5.85 with 200 ks and 20 pixels of HUBS observation, but was excluded from the candidate list due to small $\lambda_\textrm{fil}$. To the northwest of the system, there is also a background field cluster 1eRASS J121219.2+273401 ($z=0.3531$) located away from the inter-cluster connections of our interest. As a result, we modeled four clusters (three in the supercluster system and one in the background) and two filaments for our mock image.

We first generated photon lists for the filament and its connecting clusters separately. For the filament, we built a cylinder of gas with the following prescription. A rectangular box is defined with z-axis along the filament spine, and thus the x-y plane represents the filament cross-section. The box is set to be sufficiently large to enclose the filament and with uniform grids of 0.1 Mpc. Gas density and temperature for each grid were set following the filament emission model described in Sect. \ref{sec:srcmodel} along the radial direction (i.e., perpendicular to the z-axis). Gas density distribution follows the $\beta$-model in Eq. (\ref{eq:beta}) and is assumed to be constant along the z-axis. Temperature is constant at 0.5 keV for grids within an isothermal radius of 1.5 Mpc \citep{Galarraga-Espinosa2021} from the z-axis, drops log-linearly to 10$^5$ K from 1.5 Mpc to the filament radius $R_\textrm{fil}$, and remains constant at 10$^5$ K outside of $R_\textrm{fil}$. Assuming fully ionized gas and metallicity of 0.2 solar, we applied a CIE model to the grid list to generate the photon distribution. Then we projected the photons to the filament's true coordinates and inclination, and applied its corresponding Galactic absorption. 

For each of the clusters, we obtained the gas density profile by fitting a spherical $\beta$-model to multiple bands eRASS1 images using the tool MultiBand Projector 2D \citep[\texttt{MBProj2D};][]{Sanders2018}\footnote{\url{https://github.com/jeremysanders/mbproj2d}}. Details of the method are described in \citet{Liu2023} and \citet{Bulbul2024}. The cluster photons were then generated by applying a CIE model and projected to their coordinates.

The photon lists of the filament and clusters were merged and used as input for instrument simulation with \texttt{SOXS}. We added the HUBS specifications to the instrument registry, and loaded the RMF, ARF, and particle background files. We adopted the default diffuse X-ray background in \texttt{SOXS}, which contains the Galactic emission from \cite{McCammon2002}, same as the \texttt{apec}$_\textrm{MWH}$ and \texttt{apec}$_\textrm{LHB}$ used in our background model (Appendix \ref{sec:bkgmodel}). To account for CXB, a simulated point source catalog was generated using the default method in \texttt{SOXS}, which follows the flux distribution from \cite{Lehmer2012} and the spectral index distribution from \cite{Hickox2006}. The emission of line-of-sight X-ray halos was also modeled in \texttt{SOXS}, which uses a halo catalog extracted from a light cone simulation out to a redshift of 3. Finally, an event list was simulated with 200 ks HUBS exposure from the merged photon list.
\begin{figure*}
    \centering
    \includegraphics[width=0.3\linewidth]{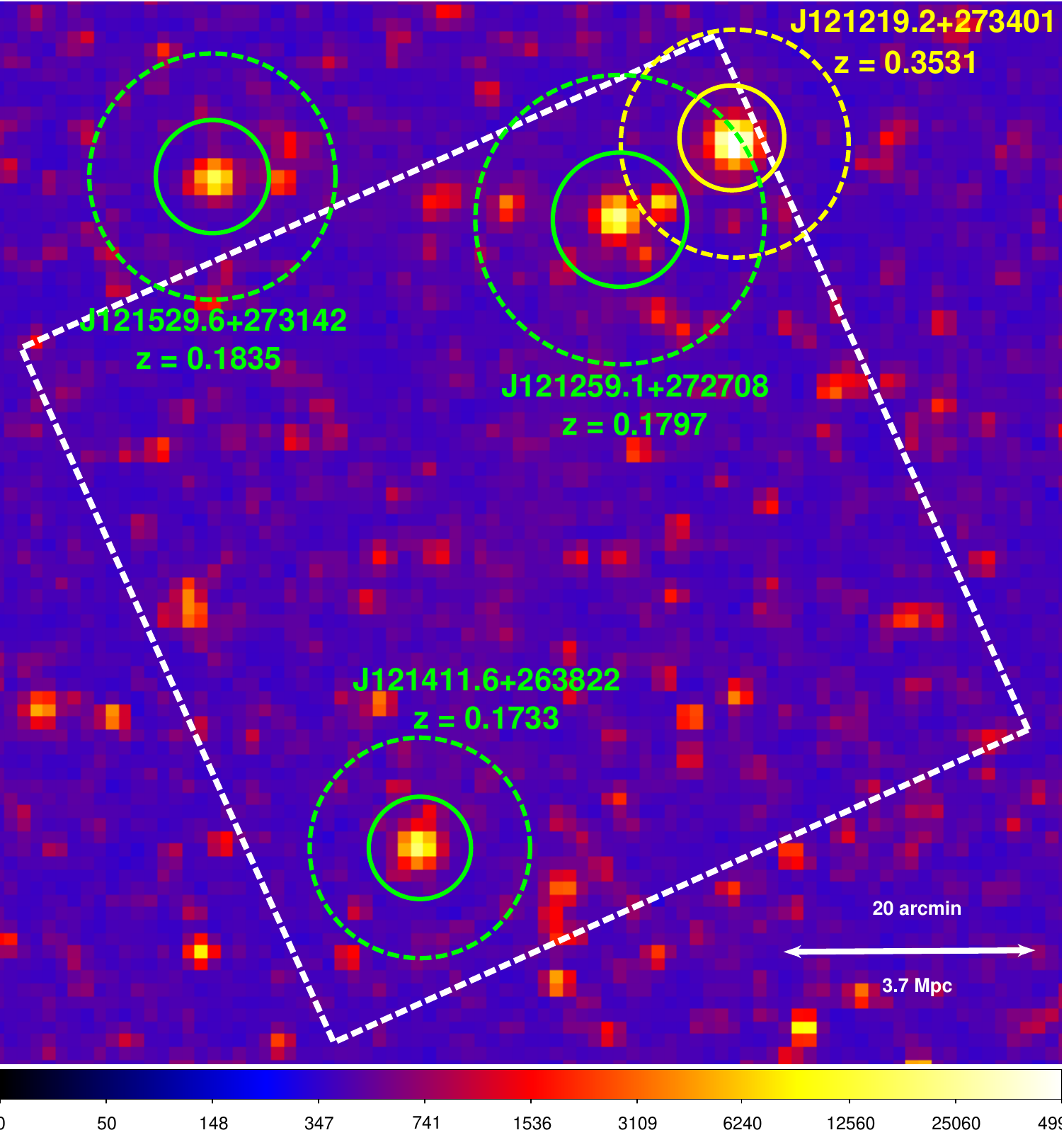}
    \includegraphics[width=0.3\linewidth]{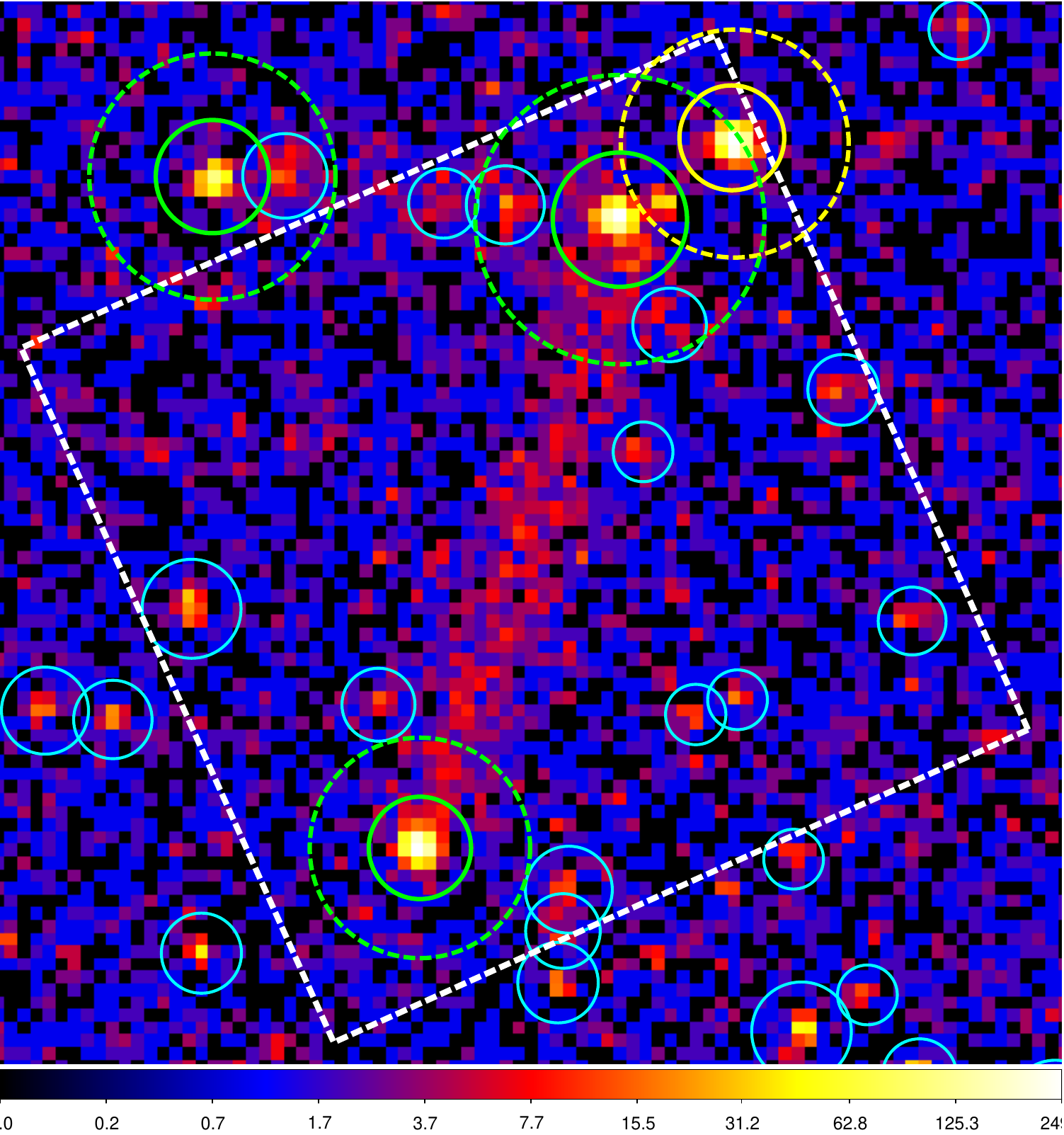}
    \includegraphics[width=0.3\linewidth]{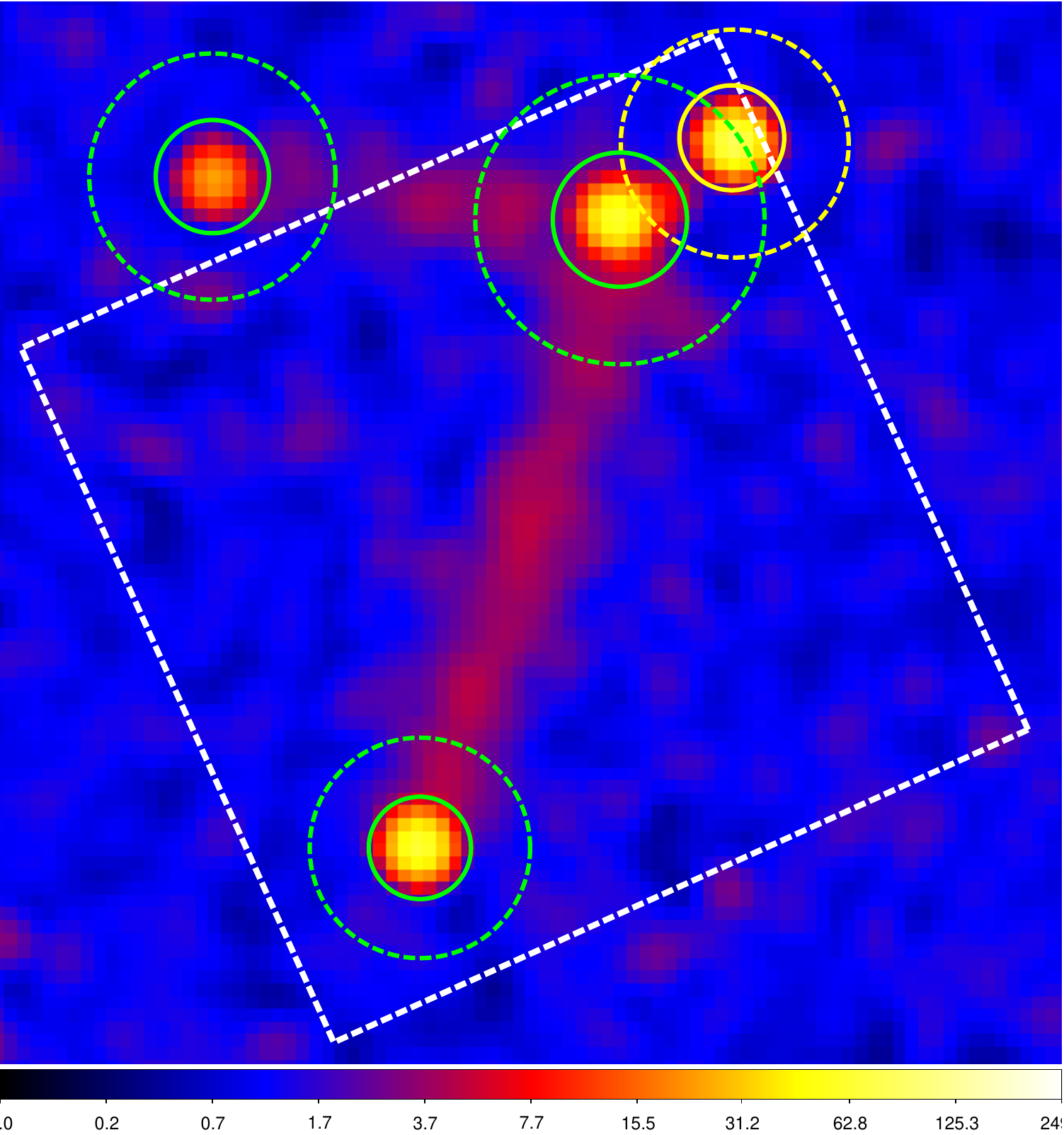}
    \caption{Mock images of 200 ks HUBS exposure for J1214+2712. \textit{Left}: counts map of the $0.1-2.0$ band; \textit{Middle}: O \textsc{viii} narrowband ($0.553-0.559$ keV) counts map; \textit{Right}: O \textsc{viii} narrowband count map with point sources masked and filled and Gaussian filter applied. Solid and dashed circles represent $R_{500}$ and $R_\textrm{vir}$ respectively for modeled galaxy clusters (green for the three clusters in the 1eRASS-SC J1214+2712 system, and yellow for the background cluster), cyan circles represent point sources, and the white dashed box in each panel represents the HUB FoV.}
    \label{fig:mockimage}
\end{figure*}
\subsubsection{Narrowband imaging}
We first created the broadband ($0.1-2.0$ keV) image from the mocked event list as shown in the left panel of Fig. \ref{fig:mockimage}, which is dominated by diffuse background emission and point sources except for the four clusters. To identify the filament emission, we filtered the events within 6 eV bandwidth around the redshift O \textsc{viii} line (in this case, $0.553-0.559$ keV) and generated a narrowband image as shown in the middle panel of Fig. \ref{fig:mockimage}, from which the filament shape can be visually identified. Furthermore, we masked out the point sources identified from the broadband image, and filled the empty pixels with a Poisson distribution of neighboring pixels, then applied a Gaussian filter with a smoothing radius of 3 pixels. Eventually, we obtained an enhanced narrowband image (right panel of Fig. \ref{fig:mockimage}) which clearly reveals the location and shape of the filament. Therefore, we conclude that HUBS is capable of directly mapping the emission of filament J1214+2712 with a single pointing of 200 ks.

\section{Discussions}\label{sec:discussions}

\subsection{Constraining chemical enrichment of filaments with HUBS}
Our mock spectra demonstrates that HUBS pointing observations will be effective not only in detecting and determining thermal properties of the filament gas, but also in constraining the metal content. 
In the $0.5-1.4$ keV range, individual elemental lines of O, Ne, Mg, and Fe can be easily distinguished (Fig. \ref{fig:mockspectrafit}). Therefore, we attempted to perform the fitting on the 200 ks mock spectra from the ``full length'' extraction region with a \texttt{TBabs} $\times$ (\texttt{vapec} + \texttt{powerlaw}) model, where the \texttt{vapec} component is used to test the constraints on individual elemental abundance, by allowing O, Ne, Mg, and Fe to vary freely, and fixing the abundance of all other elements to that of Fe (except for He, which was fixed to 1 solar). 
The results listed in Table \ref{tab:fulllength_vapec} show that: (1) the temperature and normalization constraints are at the same level as results of the \texttt{apec} fit (Sect. \ref{sec:specmock}); (2) constraints for individual elements have uncertainty of $\leq\pm0.05$ solar for most candidates (except for J0935+1949, whose upper errors reach $0.07$ solar). 

\begin{table*}
    \caption{Best-fit results for the HUBS mock spectra under the nominal case using the \texttt{vapec} model. The mock exposure time is 200 ks and the ``full length'' extraction region is used.}
    \label{tab:fulllength_vapec}

    \centering
    \begin{tabular}{c|ccccccc}
    \hline\hline
         \multirow{2}{*}{Filament Name}&  $kT$\tablefootmark{a}&  $O$\tablefootmark{a}&  $Ne$\tablefootmark{a}&  $Mg$\tablefootmark{a}& $Fe$\tablefootmark{a}& $norm$\tablefootmark{a}&$f_\textrm{SNIa error}$\tablefootmark{b}\\
 & (keV)& (solar)& (solar)& (solar)& (solar)& ($10^{-5}$ cm$^{-5}$)& \\
 \hline
         J0935+1949
&  $0.51_{-0.01}^{+0.01}$&  $0.22_{-0.01}^{+0.06}$&  $0.18_{-0.02}^{+0.05}$&  $0.24_{-0.01}^{+0.07}$&  $0.21_{-0.01}^{+0.04}$& $37.3_{-5.8}^{+1.5}$(39.3)
&$\pm 25.9\%$\\
         J0945+3418
&  

$0.50_{-0.01}^{+0.00}$&  $0.19_{-0.01}^{+0.04}$&  $0.17_{-0.01}^{+0.05}$&  $0.17_{-0.03}^{+0.03}$&  $0.19_{-0.00}^{+0.04}$& $32.4_{-5.3}^{+0.0}$(30.9)&$\pm 28.3\%$\\
         J1214+2712
&  $0.48_{-0.01}^{+0.01}$&  $0.19_{-0.01}^{+0.03}$&  $0.21_{-0.02}^{+0.03}$&  $0.20_{-0.03}^{+0.01}$&  $0.21_{-0.02}^{+0.02}$& $47.7_{-4.2}^{+4.2}$(49.1)
&$\pm 16.9\%$\\
         J2300-4534
&  

$0.49_{-0.01}^{+0.01}$&  $0.19_{-0.02}^{+0.03}$&  $0.23_{-0.02}^{+0.04}$&  $0.17_{-0.02}^{+0.04}$&  $0.22_{-0.02}^{+0.03}$& $29.6_{-3.8}^{+2.4}$(31.7)
&$\pm 24.6\%$\\
\hline
    \end{tabular}
    \tablefoot{
\\
\tablefoottext{a}{Parameters  $kT$, $O$, $Ne$, $Mg$, $Fe$ and $norm$ given in best-fit value with 68\% confidence range, followed by the model value used for spectral simulation in parenthesis.}\\
\tablefoottext{b}{$f_\textrm{SNIa error}$ is the 1$\sigma$ error in SNIa fraction obtained from \texttt{abunfit} tool using relative abundance ratios of $O$, $Ne$, $Mg$ with respect to $Fe$.}
}
\end{table*}

The metallicity of individual elements can be used to trace the chemical evolution of the filament gas since few stellar processes are responsible for the generation of metals. The elements of interest in this work, namely O, Ne, Mg, and Fe, are produced and dissipated through different types of supernova explosions. The $\alpha$ elements such as O, Ne, and Mg are products of core-collapse supernovae (SNcc), and heavier elements such as Fe are products of type Ia supernovae (SNIa). Thus, it is possible to estimate the contributions of SNcc and SNIa to the chemical enrichment of WHIM in cosmic filaments by fitting elemental abundance ratios.

We used the \texttt{abunfit}\footnote{\url{https://github.com/mernier/abunfit/}} tool developed by \cite{Mernier2016}, which allows for fitting of elemental abundance ratios with respect to Fe (X/Fe) with a combination of supernovae yield models. We adopt the 3D N40 model from \cite{Seitenzahl2013} for SNIa, \cite{Nomoto2013} model with initial metallicity of zero for SNcc, and a Salpeter initial mass function \citep{Salpeter1955}. The contribution fractions of SNIa and SNcc can be fitted with X/Fe values. The \texttt{vapec} fitting of our leading candidate J1214+2712 results in O/Fe, Mg/Fe, and Ne/Fe ratios with 1$\sigma$ uncertainty of $\sim\pm0.15$, which leads to the best-fit SNIa ratio $f_\textrm{SNIa}$, the relative number of SNIa over the total number of supernova events (i.e., $f_\textrm{SNIa} = \textrm{N}_\textrm{SNIa} / (\textrm{N}_\textrm{SNIa} + \textrm{N}_\textrm{SNcc})$), to have 1$\sigma$ uncertainty of $\pm16.9\%$. This precision, in combination with the future elemental abundance measurement in galaxy cluster outskirts using HUBS, is crucial in tracing the chemical origin of filaments and galaxy clusters.

\subsection{Conservative assumptions for gas properties}
The nominal gas properties assumed in this work, as suggested by previous soft X-ray observations through various techniques, represent the hotter, denser end of the X-ray emitting WHIM phase gas studied in numerical simulations \citep{Tanimura2022, Veronica2024, Zhang2024, Dietl2024}. 
Therefore, it is also necessary to test more conservative cases, in which WHIM gas phase is bounded by a lower temperature and lower density than those found in observations. For this purpose, we considered an extreme scenario in which all the WHIM gas has an overall temperature of $kT=0.1$ keV (the lower boundary of gas to be traced by X-ray emission, see Fig. 10 in \citealt{Zhang2024}). At the same time, we reduced the gas density to half of the nominal case ($\delta_0=100$), which translates to normalization parameter of a quarter of the nominal case, to make the conservative case truly representative of a worst-case scenario.

We followed the same procedure in Sect. \ref{sec:specmock} to generate mock spectra for the ``full length'' region. With the reduced gas temperature and density, the source continuum becomes almost completely overwhelmed by the background, and the only source emission lines with intensity comparable to the background are the redshifted O \textsc{vii}\textit{(r)} and O \textsc{vii}\textit{(f)} (rest-frame energies are 0.574 keV and 0.561 keV respectively). Therefore we chose the rest-frame energy range of $0.2-0.6$ keV to fit the \texttt{TBabs} $\times$ (\texttt{apec} + \texttt{powerlaw}) model. Since the 200 ks exposure wasn't sufficient to obtain enough source counts for spectral fitting, we tried to increase exposure time for the mock spectra until the net counts within the fitting energy could consistently reach more than 1000 for all the candidates, and eventually found that 1 Ms of exposure is required. Fitting results for the 1 Ms mock spectra of the conservative case are listed in Table \ref{tab:worstcase_fit}.

Due to severe density-metallicity degeneracy, we had to fix the metallicity parameter at the simulation input value of 0.2 solar, which then allowed model values of temperature and normalization to be recovered within 68\% uncertainties. The derived gas density offset for all candidates are within $10\%$, and 68\% errors are within $\pm20\%$ of the best-fit gas density results for all candidates. 

Furthermore, we evaluated the error in determining gas density if a wrong metallicity is assumed under this conservative scenario. To mimic the situation, we used metallicity of 0.1 solar when simulating the mock spectra, while still fixing it at 0.2 solar during spectral fitting. As shown in Table \ref{tab:worstcase_fit_wrongZ}, due to falsely assumed metallicity, the density-metallicity degeneracy will cause a $30-40\%$ offset and up to $\sim50\%$ error in the gas density measurement. 

Our tests for the conservative gas properties show that, even in the unlikely case that filaments completely composed of lower temperature and lower density gas than those implied from current observations, acceptable constraints on gas density can still be obtained with 1 Ms of HUBS pointing exposure for each candidate filaments using spectra extracted from the ``full length'' region. However, if metallicity of the filaments is falsely assumed, a significant offset in gas mass measurement can occur.

\begin{table*}
    \caption{Best-fit results for the HUBS mock spectra under the conservative case using the \texttt{apec} model. The mock exposure time is 1 Ms and the ``full length'' extraction region is used.}
    \label{tab:worstcase_fit}
   
   \centering
    \begin{tabular}{cccccccc}
        \hline\hline
         \multirow{2}{*}{Filament Name}&  $kT$\tablefootmark{a}&  $Z$\tablefootmark{b}&  $norm$\tablefootmark{a}&C-stat/dof  & Net counts\tablefootmark{c}& $\delta_\textrm{offset}$ \tablefootmark{d}&$\delta_\textrm{error}$\tablefootmark{d}\\
 & (keV)& (solar)& ($10^{-5}$ cm$^{-5}$)& ($10^{-5}$ cm$^{-5}$)&  & &\\
 \hline
         J0935+1949
&  $0.10_{-0.01}^{+0.01}$(0.1)&  0.2f(0.2)&  $10.5_{-3.2}^{+4.0}$(9.8)&257.8/270& 1379& 3.4\%&$_{-16.5\%}^{+17.4\%}$
\\
         J0945+3418
&  $0.10_{-0.01}^{+0.01}$(0.1)&  0.2f(0.2)&  $7.8_{-1.5}^{+1.1}$(7.7)&295.5/283& 2566& 0.4\%&$_{-10.4\%}^{+7.0\%}$
\\
         J1214+2712
&  $0.10_{-0.00}^{+0.00}$(0.1)&  0.2f(0.2)&  $13.2_{-2.2}^{+2.9}$(12.3)&259.3/274& 2548& 3.7\%&$_{-8.9\%}^{+10.5\%}$
\\
 J2300-4534
& $0.10_{-0.00}^{+0.00}$(0.1)& 0.2f(0.2)& $9.6_{-1.4}^{+0.9}$(7.9)& 287.4/282& 1903& 9.9\%&$_{-7.5\%}^{+4.8\%}$
\\
 \hline
    \end{tabular}
\tablefoot{
\\
\tablefoottext{a}{Parameters $kT$ and $norm$ given in best-fit value with 68\% confidence range, followed by the model value used for spectral simulation in parenthesis.}\\
\tablefoottext{b}{Parameters $Z$ given in fixed value followed by the model value used for spectral simulation in parenthesis.}\\
\tablefoottext{c}{Net counts of the spectra in the $0.2-0.6$ keV rest-frame energy range after background subtraction.}\\
\tablefoottext{d}{$\delta_\textrm{offset}$ and $\delta_\textrm{error}$ are the deviation from simulation input value and 68\% error of the best-fit value of the gas density inferred from $norm$ fitting results following Eq. (\ref{eq:norm}) respectively.}
}
\end{table*}

\begin{table*}
    \caption{Best-fit results for the HUBS mock spectra under the conservative case with a falsely assumed metallicity value. The mock exposure time is 1 Ms and the ``full length'' extraction region is used. }
    \label{tab:worstcase_fit_wrongZ}
   \centering
   
    \begin{tabular}{cccccccc}
        \hline\hline
         \multirow{2}{*}{Filament Name}&  $kT$\tablefootmark{a}&  $Z$\tablefootmark{b}&  $norm$\tablefootmark{a}&C-stat/dof  & Net counts\tablefootmark{c}& $\delta_\textrm{offset}$ \tablefootmark{d}&$\delta_\textrm{error}$\tablefootmark{d}\\
 & (keV)& (solar)& ($10^{-5}$ cm$^{-5}$)& ($10^{-5}$ cm$^{-5}$)&  & &\\
 \hline
         J0935+1949
&  $0.11_{-0.02}^{+0.06}$(0.1)&  0.2f(0.1)&  $3.7_{-2.0}^{+2.6}$(9.8)&275.1/270& 354& 38.3\%&$_{-32.0\%}^{+29.8\%}$
\\
         J0945+3418
&  $0.10_{-0.01}^{+0.11}$(0.1)&  0.2f(0.1)&  $2.9_{-1.7}^{+1.2}$(7.7)&274.6/282& 962& 38.9\%&$_{-36.3\%}^{+18.7\%}$
\\
         J1214+2712
&  $0.10_{-0.02}^{+0.02}$(0.1)&  0.2f(0.1)&  $4.6_{-2.7}^{+2.8}$(12.3)&240.4/273& 796& 38.5\%&$_{-34.6\%}^{+26.2\%}$
\\
 J2300-4534
& $0.10_{-0.01}^{+0.01}$(0.1)& 0.2f(0.1)& $3.7_{-1.4}^{+0.8}$(7.9)& 250.6/283& 1084& 31.9\%&$_{-20.9\%}^{+10.1\%}$
\\
 \hline
    \end{tabular}
\tablefoot{
\\
\tablefoottext{a}{Parameters $kT$ and $norm$ given in best-fit value with 68\% confidence range, followed by the model value used for spectral simulation in parenthesis.}\\
\tablefoottext{b}{Parameters $Z$ given in fixed value followed by the model value used for spectral simulation in parenthesis.}\\
\tablefoottext{c}{Net counts of the spectra in the $0.2-0.6$ keV rest-frame energy range after background subtraction.}\\
\tablefoottext{d}{$\delta_\textrm{offset}$ and $\delta_\textrm{error}$ are the deviation from simulation input value and 68\% error of the best-fit value of the gas density inferred from $norm$ fitting results following Eq. (\ref{eq:norm}) respectively.}
}
\end{table*}

\subsection{Deviation from CIE}
In this paper we approximated the WHIM gas to be in CIE condition, which was also assumed in recent observational as well as other mock observation works \citep[e.g.,][]{Tanimura2020, Ghirardini2021, Gouin2023, Veronica2024, Zhang2024, Dietl2024}. 
However, photoionization due to the cosmic ultra-violet and X-ray background in addition to collisional ionization may significantly impact the cooling rate and ionization fractions of metals \citep{Wiersma2009}. Additionally, the low density of WHIM may result in ionization and recombination timescales comparable to the age of the Universe \citep{Yoshikawa2006}, causing the ionization state of metals to deviate from ionization equilibrium. 
Since future detections and diagnostics of WHIM with high-resolution spectroscopy is highly dependent on metal emission lines, we need to evaluate the deviation from CIE for the gas density and temperature in the core of filaments modeled in our analysis. Specifically, we evaluate the emission lines of O \textsc{vii} and O \textsc{viii} ions, which can be best characterized with HUBS. 

First of all, a simple comparison can be made between the photoionization and the collisional ionization rates following \cite{Wong2011}, who estimated UV photoionization rate at $z=0$ to be $5.2 \times 10^{-18}\ \textrm{s}^{-1}$ and $1.7 \times 10^{-18}\ \textrm{s}^{-1}$ for O \textsc{vii} and O \textsc{viii} respectively for the outskirts of galaxy clusters, where a similar gas density to that of cosmic filaments can be found. On the other hand, the slowest collision ionization rate (1/$n_\textrm{e}t$) of oxygen ions for temperatures within $0.1-0.5$ keV is more than $1-2\ \times\ 10^{-12}\ \textrm{cm}^{3}\ \textrm{s}^{-1}$ (see Fig. 1 in \citealt{Smith2010}), which translates to $\sim5 \times 10^{-17}\ \textrm{s}^{-1}$ for $n_\textrm{e} \approx 3-4\ \times 10^{-5}\ \textrm{cm}^{-3}$ (average core density for candidate filaments), at least one order of magnitude higher than the photoionization rate. Therefore, we can neglect the photoionization effect for the core of the filaments, which is minor compared with collisional ionization.

To evaluate the effect of non-equilibrium ionization, we compared the time required to reach equilibrium for gas with density and temperature of our filament cores with the time passed after the shock. We assume a simple thermal history in which the majority of shock heating happened before $z=1$, since several numerical simulations have found little evolution of filament density since $z\sim1$ \citep{Cui2019, Galarraga-Espinosa2024}, indicating the core of filaments has been formed $\sim 5$ Gyrs ago. Again based on results from \cite{Smith2010}, the maximum density-weighted equilibrium timescale for oxygen is $\sim 10^{12}\ \textrm{cm}^{-3}\ \textrm{s}$ for plasma temperature of $\sim 10^6\ \textrm{K}$. For the core of filaments with an average density of $n_\textrm{e} \approx 3-4\ \times 10^{-5}\ \textrm{cm}^{-3}$, the time required to reach equilibrium is $\sim 1$ Gyr. Therefore, even with low density, gas in the filament core would have enough time to converge to equilibrium after the major shocks.

Furthermore, \cite{Yoshikawa2006} performed not only calculations of a simple thermal history (shock at $z=1$) but also cosmological hydrodynamical simulations considering both photo- and collisional ionization processes. As a result, all of the O \textsc{vii}, O \textsc{viii}, O \textsc{ix} ions with overdensity larger than 100 and temperature larger than $10^{6.5}$ K are close to ionization equilibrium. However, for gas with overdensity of around 10 and temperature around $10^6$ K, there is a significant increase in the fraction of O \textsc{vii} and a decrease in the fraction of O \textsc{ix}.

Overall, these results support our approximation of the CIE condition for the filament core under our typical setting of gas properties. However, if the conservative properties are adopted, a boost in the O \textsc{vii} line is expected, which is favorable for narrowband detection, but will pose challenges for spectral diagnostics for having different line ratios between oxygen species than the CIE condition.

\subsection{Mass of clusters connected to candidate filaments}
 
A number of recent numerical and observational works \citep[e.g.][]{DarraghFord2019, Sarron2019,Gouin2021, Angelinelli2021, Santoni2024} have found that more massive clusters tend to have higher connectivity (i.e., the number of filaments connected to them). Therefore, one may find it desirable to select clusters with higher mass. However, very few studies provided quantitative analysis on the correlation between filament gas properties and cluster mass to determine a concrete mass threshold. 

\cite{Angelinelli2021} compared simulated gas properties of filaments around 13 galaxy clusters in the $5\times10^{13} \leq M_{100}/M_\odot \leq 4\times 10^{14}$ mass range and found that the temperature of the filaments slightly increases with the mass of the cluster, while the gas density of filaments appears to be independent of the mass of the cluster. \cite{Ilc2024} performed zoom-in simulations around five clusters in the Dianoga simulations, and found that the filaments around the less massive clusters actually have a higher density close to their spine, but the density falls to a similar value as those around more massive clusters at larger radii. 

In observations, while filaments have been detected in the outskirts of the most massive clusters, for example Abell 2744 with $M_{200} = 2.22_{-0.12}^{+0.13}\times 10^{15} M_\odot$ \citep{Babyk2012} and Abell 3667 with $M_{500} = 7.04\pm 0.05 \times10^{14} M_\odot$ \citep{deGasperin2022}, they were also detected around less massive systems such as Abell 3391 with $M_{500} = 2.16\times10^{14} M_\odot$ \citep{Alvarez2018}, Abell 1750N with $M_{500} = 1.98 \times 10^{14} M_\odot$ \citep{Bulbul2016}, and Abell 98 with $M_{500} \sim2 \times
10^{14} M_\odot$ \citep{Paterno-Mahler2014}.

We notice that the majority of clusters connected to our selected filament candidates have $M_{500}$ of $1-2 \times10^{14} M_\odot$, which are not among the most massive galaxy clusters. We reviewed our selection procedure and identified two additional filaments which are connected to at least one massive cluster, but were eliminated because of the small projected length $\lambda_\textrm{fil}$ (Table \ref{tab:massiveclusterproperties}). They may provide a valuable addition to the HUBS observation targets, given that their cluster emission and clumps within the filaments can be accurately modeled through extended exposure times and multi-wavelength observations.

\begin{table*}
\caption{Properties of additional candidate filaments connected to massive clusters.}
\label{tab:massiveclusterproperties}
    \centering
    \begin{tabular}{c|ccc|ccc|c}
        \hline 
        \hline

        \multirow{2}{*}{Filament Name}&\multirow{2}{*}{Cluster Name}& $z$&$M_{500}$& $D_\textrm{phys}$ & $\lambda_\textrm{fil}$&$\theta$ & O \textsc{viii} $S/N$\\
        & & (-)& ($10^{14} M_\odot$)& (Mpc) & (arcmin)&(deg) & (-)\\
        \hline
        \multirow{2}{*}{J0859+0306}&1eRASS J085751.0+031014&$0.2025\pm0.0012$&$6.20_{-0.67}^{+0.60}$&\multirow{2}{*}{40.3
}&\multirow{2}{*}{4.21
}&\multirow{2}{*}{7.4
}&\multirow{2}{*}{8.71
}\\
        &1eRASS J085932.4+030832&$0.1955\pm0.0010$&$1.84_{-0.50}^{+0.41}$& & & & \\
        \hline
        \multirow{2}{*}{J1242+2731}&1eRASS J124325.7+271700&$0.1924\pm0.0022$&$5.83_{-0.65}^{+0.46}$&\multirow{2}{*}{20.5
}&\multirow{2}{*}{3.88
}&\multirow{2}{*}{14.7}&\multirow{2}{*}{6.39}\\
        &1eRASS J124356.3+274155&$0.1958\pm0.0013$&$2.19_{-0.34}^{+0.27}$& & & & \\
\hline
    \end{tabular}
\tablefoot{Member clusters properties, geometrical properties, and O \textsc{viii} $S/N$ of the two additional candidate filaments connected to massive clusters.}
\end{table*}

\section{Conclusion and outlook}\label{sec:concl}

In this work, we constructed a sample of inter-cluster filaments based on the eRASS1 supercluster catalog \citep{Liu2024} and built a model to estimate their soft X-ray emission. From this sample, we selected four candidates appropriate for future pointing observations with the next-generation HUBS mission. For these candidates, we simulated and analyzed mock spectra and images using HUBS instrumental specifications with 200 ks of exposure time, and found the following conclusions:
\begin{itemize}
    \item WHIM gas properties including temperature, metallicity, and density can be accurately recovered with 200 ks HUBS exposure on each of the candidate filaments. The expected 68\% uncertainties ($\pm0.01$ keV for temperature, $\leq\pm0.03$ solar for metallicity, and $<\pm10\%$ for density) will be a significant improvement of the current instruments, and will place unprecedented constraints on the cosmic baryon and metal budget.
    \item Spatial variation of gas density of $\sim\pm20\%$ can be distinguished on the moderate resolution of 20 arcmin$^2$, which allows for segmentation of $8-10$ portions of each candidate filament. 
    \item The abundance of individual elements including O, Ne, Mg, and Fe can be accurately determined with 68\% uncertainties of $\leq\pm0.05$ solar for most candidates, which will enable understanding of the chemical origin by fitting the contributions from different enrichment processes (i.e., supernovae). 
    \item Narrowband imaging provides a means for directly mapping the spatial distribution of WHIM. We showed with O \textsc{viii} mapping that it can clearly reveal the location and shape of our filaments.
    \item Even with a conservative assumption of gas properties (i.e., lower temperature and lower density), the WHIM mass in filaments can still be measured with acceptable uncertainty if the exposure time is increased to 1 Ms. It is, however, crucial to assume the correct metallicity when analyzing the cooler gas.
    \item Two additional filaments were identified, which are connected to at least one massive cluster with $M_{500} > 5 \times10^{14} M_\odot$, but fail to meet the criteria for projected angular size. They may be suitable targets if contamination of cluster and clump emission can be accurately modeled and separated from filament emission. 
\end{itemize}

Using mock observations conducted with HUBS, our work exemplifies the substantial improvements in spectral characterization of the WHIM in cosmic filaments that future missions with advanced spectroscopic capabilities will bring. Furthermore, future instruments such as LEM or Athena X-IFU, with more ambitious specifications in terms of both effective area in the soft band and spatial resolution, will be complimentary to the large FoV of HUBS and bring forth more precise measurements to solve the longstanding mysteries of baryon census, structure formation, and chemical evolution.
\section*{Data availability}
The full Table B.1 is only available in electronic form at the CDS via anonymous ftp to \url{cdsarc.u-strasbg.fr} (\url{130.79.128.5}) or via \url{http://cdsweb.u-strasbg.fr/cgi-bin/qcat?J/A+A/}.

\begin{acknowledgements}

We thank the anonymous referee for the thoughtful comments that improved the paper.
We thank Dr. Xianzhong Zheng, Dr. Weiwei Xu, Dr. Thomas Reiprich, Dr. Ming Sun, Mr. Hoongwah Siew, Mr. Cheng Liu, and Mr. Chenxi Shan for their valuable feedback. This work was supported by the National Natural Science Foundation of China (NFSC) Grant No. 12233005. YZ acknowledges the financial support of the National Center for High-Level Talent Training in Mathematics, Physics, Chemistry, and Biology. DH and NW acknowledge the financial support of the GA\v{C}R EXPRO grant No. 21-13491X.
\end{acknowledgements}

%
%
\bibliographystyle{aa}
\bibliography{ref}

\begin{appendix}

\section{Background modeling}\label{sec:bkgmodel}
When estimating the emission line $S/N$ of filaments in our sample and generating mock observations for our candidates, a background model was generated with the following prescription. The total background expected for HUBS observations consists of an astrophysical background in the soft X-ray band and a particle background caused by the interaction between incident charged particles and the detector. 

\subsection{Astrophysical background}
The astrophysical background is comprised of foreground emission due to the Milky Way and background emission due to cosmic sources. 
The Galactic foreground is contributed by continuum and line emission of thermal hot gas in and around the Milky Way \citep{Kuntz2000}, typically modeled as that of a $\sim0.2$ keV gas in the Milky Way Halo (MWH) absorbed by the ISM \citep{Henley2013}, and a $\sim0.1$ keV gas in the Local Hot Bubble (LHB) \citep{Liu2017}. Recently, a $\sim0.8$ keV component was discovered at lower Galactic latitudes \citep{Sugiyama2023}. On the other hand, the cosmic X-ray background (CXB) is contributed by unresolved point sources (most of which are AGNs) \citep{Mushotzky2000}.

We followed the prescription given in \cite{Lotti2014} who modeled the in-orbit background for the Athena mission to evaluate its effects on observations. The model parameters were taken from \cite{McCammon2002}, who extracted the diffuse soft X-ray background in the 60 - 1000 eV band using a microcalorimeter onboard a sounding rocket pointed towards a 1 sr region centered at $l = 90^{\circ}$, $b = 60^{\circ}$, representative of high Galactic latitude observations. The model is written as
\begin{equation}\label{eq:bkg}
    \texttt{Model}_\textrm{bkg} =\ \texttt{TBabs}\ \times\ (\texttt{powerlaw}\ + \ \texttt{apec}_\textrm{MWH})\ +\ \texttt{apec}_\textrm{LHB},
\end{equation}
where \texttt{TBabs} represents photo-absorption caused by the ISM, \texttt{powerlaw} represents composite emission of the CXB, $\texttt{apec}_\textrm{MWH}$ represents emission of the MWH, and $\texttt{apec}_\textrm{LHB}$ represents emission of the LHB, with model parameters listed in Table \ref{tab:sky_bkg}. It is worth noting that in \cite{Lotti2014}, the authors removed 80\% of the CXB emission based on the assumption that Athena can resolve the majority of bright point sources. We conservatively made no such assumption for HUBS; thus, the \texttt{powerlaw} $norm$ we used is five times of that in \cite{Lotti2014}. 

\begin{table}
    \caption{Parameters for the astrophysical background model $\texttt{Model}_\textrm{bkg}$.}
    \label{tab:sky_bkg}
    \centering
    \begin{tabular}{cccc}
        \hline
        \hline
         Component&Parameter&Unit& Value\\
        \hline
         \multirow{2}{*}{\texttt{powerlaw}$_\textrm{CXB}$}&  $\Gamma$&-& 1.52\\
         &$norm$&keV$^{-1}$ cm$^{-2}$ s$^{-1}$& 1.0e-6\\
        \hline
         \multirow{4}{*}{\texttt{apec}$_\textrm{MWH}$}&  $kT$&keV& 0.225\\
         &$Z$&solar&1\\
         &$z$&-&0.0\\
         &$norm$&$\textrm{cm}^{-5}$& 7.3e-7\\
        \hline
         \multirow{4}{*}{\texttt{apec}$_\textrm{LHB}$}& $kT$&keV&0.099\\
         &$Z$& solar&1\\
         &$z$&-&0.0\\
         &$norm$&$\textrm{cm}^{-5}$&1.7e-6\\  
        \hline    
    \end{tabular}
    \tablefoot{Galactic and CXB model parameters were adopted from  \cite{McCammon2002}, normalized to 1 arcmin$^2$.}
\end{table}

\subsection{Particle background} 
Same as all X-ray satellites, HUBS will suffer from a particle background (also referred to as non-X-ray background; NXB) mainly caused by charged particles traveling in space and interacting with the detector. After the particle energy is deposited on the detector, contamination counts are generated, which are difficult to distinguish from the X-ray photons. Thus, we reference the simulation results from XRISM which is also a microcalorimeter mission operating in the low-earth orbit. By adopting the upper limit of the simulated XRISM particle background of 2e-3 counts s$^{-1}$ keV$^{-1}$ for the array \citep{XRISMScienceTeam2022}, we derived an equivalent value based on HUBS detector size and focal length, which is converted to HUBS particle background level of 3e-5 counts s$^{-1}$ keV$^{-1}$ arcmin$^{-2}$.

\subsection{Mock background spectra}
\begin{figure}
    \centering
    \includegraphics[width=0.9\hsize]{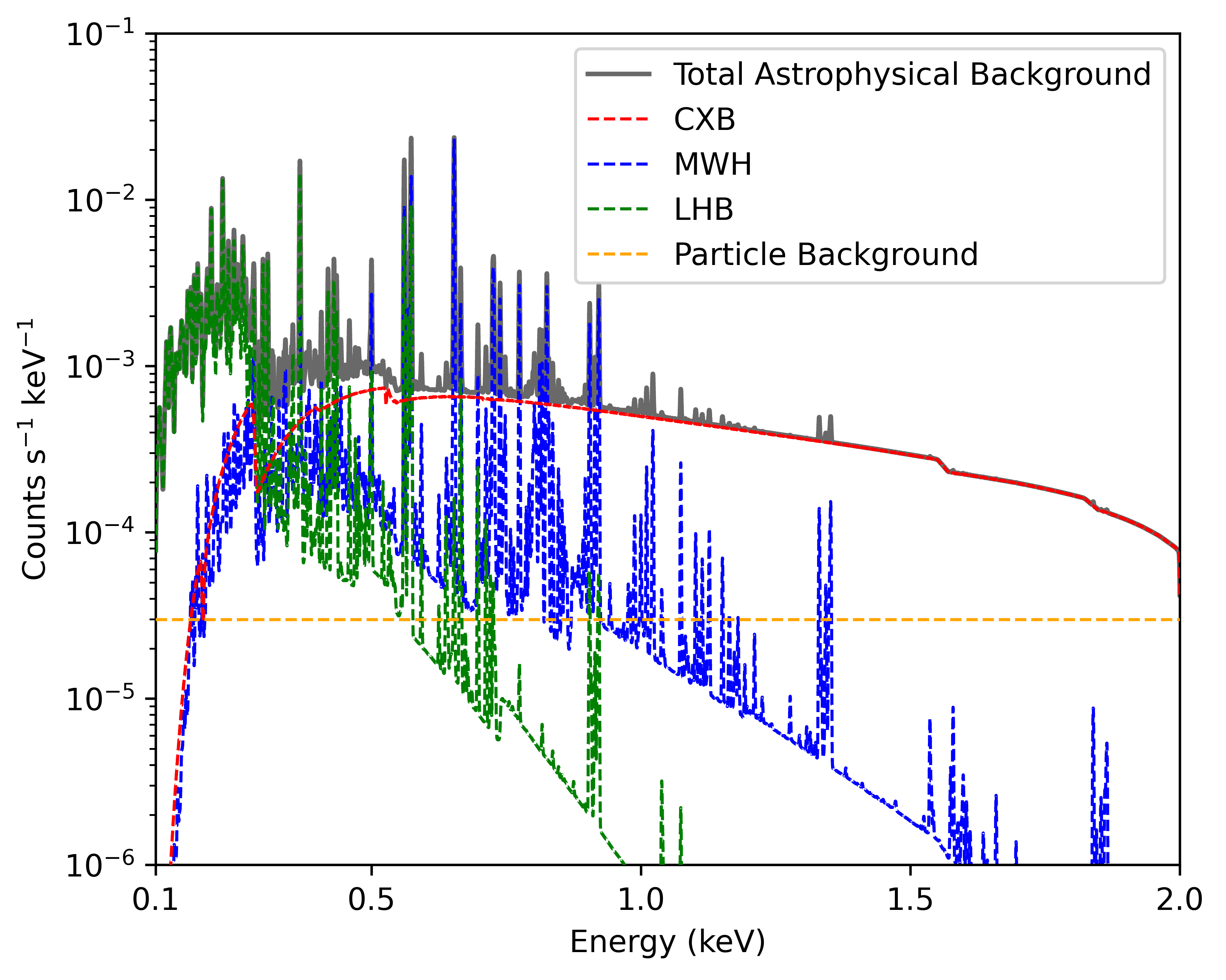}
    \caption{Simulated astrophysical and particle background spectra. The total astrophysical background (grey solid line) is the sum of the CXB (red dashed line), MWH (blue dashed line), and LHB (green dashed line). The particle background level is indicated by the orange dashed line.}
    \label{fig:background_components}
\end{figure}

Using the \texttt{XSPEC} tool \texttt{fakeit} and HUBS response and effective area files, we simulated the background discussed above (Fig. \ref{fig:background_components}), where the $0.1-2.0$ keV HUBS energy range can be divided into a soft band ($0.1-0.5$ keV) and a hard band ($0.5-2.0$ keV). We find that in the hard band, the background continuum is dominated by the CXB component while emission lines are mostly contributed by metal emission of the MHW component. In the soft band, on the other hand, the background is dominated by the LHB component. Throughout the entire energy range, the astrophysical background exceeds the particle background by more than an order of magnitude. Therefore, particle background is ignored when calculating the O \textsc{viii} $S/N$ (Sect. \ref{sec:SNR}) and simulating the mock spectra (Sect. \ref{sec:specmock}), but considered when generating the mock image (Sect. \ref{sec:SOXSmodel}).

\section{Filament sample properties}
\begin{figure}[h!]
    \centering
    
    \includegraphics[width=0.36\textwidth]{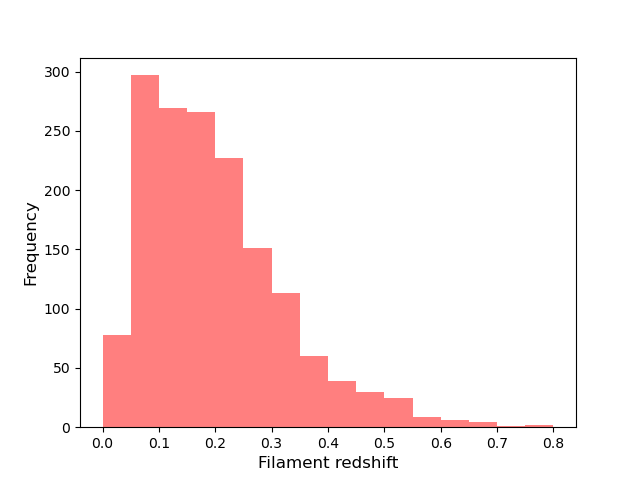}
    \includegraphics[width=0.36\textwidth]{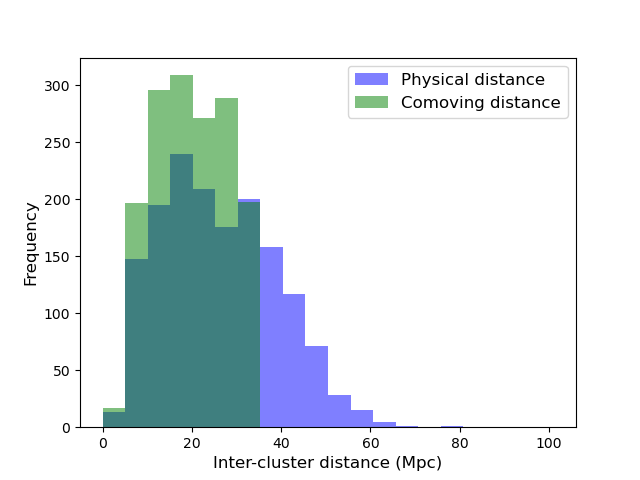}
    \includegraphics[width=0.36\textwidth]{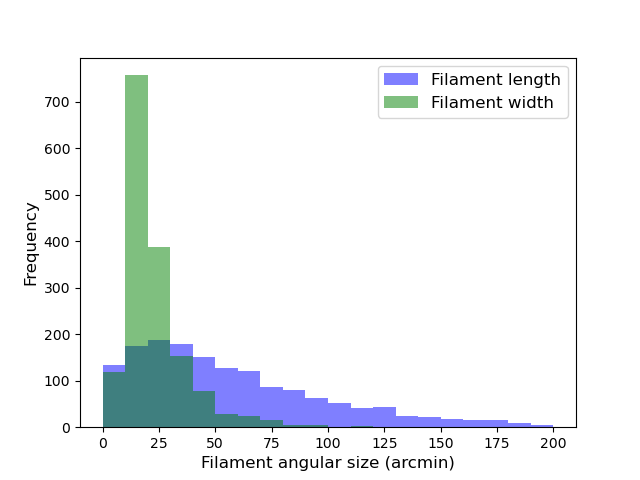}
    \includegraphics[width=0.36\textwidth]{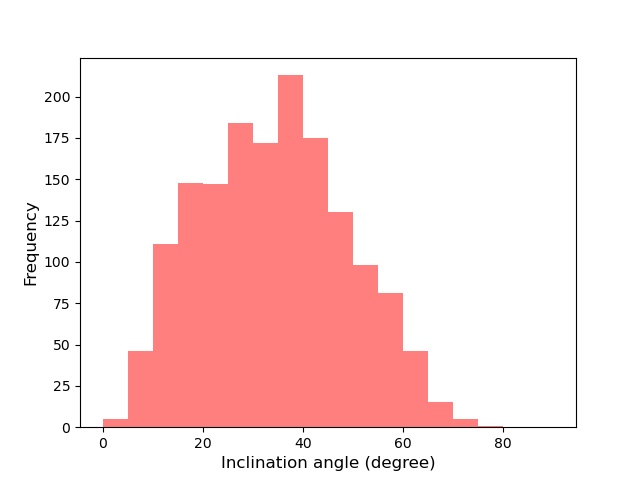}
    \caption{Distribution of filament properties.}
    \label{fig:filamentStats}
\end{figure}

A number of physical and emission properties were calculated and extracted in Sect. \ref{sec:geometry}, \ref{sec:srcmodel}, and \ref{sec:select}. The distribution of redshift, inter-cluster separation, angular size, and inclination angle are plotted in Fig. \ref{fig:filamentStats}. We list in Table \ref{tab:sampleproperties} the relevant properties for all 1577 filaments in the sample. The full table can be accessed through the CDS.
\begin{sidewaystable}
    \caption{Physical and emission properties of the inter-cluster filament sample.}
    \centering
    
    \begin{tabular}{cc|cc|ccccccc|c}
    \hline
    \hline
           \multirow{2}{*}{GC1 name}&  \multirow{2}{*}{GC2 name}&  $z_\textrm{avg}$&  $N_\textrm{H}$&   $D_\textrm{C}$&$D_\textrm{phys}$&  $\delta_\textrm{proj}$& $\omega_\textrm{fil}$& $\omega_\textrm{core}$&$\lambda_\textrm{fil}$ & $\theta$&O \textsc{viii} $S/N$\\
           &  &  (-)&  ($10^{20}$ cm$^{-2}$)&   ($h^{-1}$ Mpc)&(Mpc) &  (deg)& (arcmin)& (arcmin)& (arcmin)& (deg)&(-)\\
         \hline
           1eRASS J000236.8-471113&  1eRASS J000104.0-464435
&  0.17105
&  1.48 
&   10.2 
&13.0 
&  0.52 
& 18.09 
& 6.18 
& 13.08 
& 25.7 
&2.02
\\
           1eRASS J000203.6-481447&  1eRASS J000236.8-471113
&  0.1668
&  1.34 
&   30.0 
&39.3 
&  1.07 
& 16.07 
& 6.31 
& 47.85 
& 16.8 
&2.86
\\
           1eRASS J235556.4-511029&  1eRASS J000654.8-503210
&  0.11895
&  1.46 
&   17.4 
&19.7 
&  1.85 
& 24.41 
& 8.38 
& 86.48 
& 48.6 
&2.64
\\
           1eRASS J000607.8-513332&  1eRASS J000654.8-503210
&  0.12225
&  1.48 
&   24.2 
&29.7 
&  1.05 
& 22.14 
& 8.19 
& 40.53 
& 16.8 
&5.22
\\
           1eRASS J000425.6-561726&  1eRASS J000155.6-560812
&  0.29405
&  1.36 
&   21.0 
&33.0 
&  0.38 
& 14.35 
& 4.10 
& 8.46 
& 10.9 
&8.77
\\
  1eRASS J000557.9-562841& 1eRASS J000155.6-560812
& 0.29135
& 1.37 
&  13.8 
&17.9 
& 0.66 
& 15.55 
& 4.13 
& 23.79 
& 36.6 
&4.14
\\
 1eRASS J001250.4-691851& 1eRASS J000056.7-682010
& 0.24295
& 2.96 
& 28.3 
& 36.4 
& 1.46 
& 13.18 
& 4.71 
& 74.39 
& 34.9 
&3.83
\\
 1eRASS J000056.7-682010& 1eRASS J234947.9-671802
& 0.2421
& 2.83 
& 26.5 
& 33.2 
& 1.48 
& 13.05 
& 4.72 
& 75.52 
& 39.3 
&3.47
\\
           ...&  &  &  &   &&  & & & & &\\
         \hline
    \end{tabular}
\tablefoot{
$z_\textrm{avg}$ and $N_\textrm{H}$ were obtained following Sect. \ref{sec:srcmodel}, $D_\textrm{C}$ is the comoving distance, $D_\textrm{phys}$, $\delta_\textrm{proj}$, $\omega_\textrm{fil}$, $\lambda_\textrm{fil}$, and $\theta$ were described in Fig. \ref{fig:geometry}, $\omega_\textrm{core}$ is the angular size corresponding to the core radius $r_\textrm{c}=0.56$ Mpc, and O \textsc{viii} $S/N$ was calculated based on Sect. \ref{sec:SNR}. The full table is available at the CDS.
}
    \label{tab:sampleproperties}
\end{sidewaystable}

\end{appendix}
\end{document}